\documentclass[nohyper]{JHEP3}

\usepackage{graphics}

\newif\ifdraft
\drafttrue
\newif\ifpreprint
\preprinttrue

\def\beq{\begin{equation}}
\def\eeq{\end{equation}}
\def\beqa{\begin{eqnarray}}
\def\eeqa{\end{eqnarray}}

\def \Tr {\mbox{Tr\,}}

\newcommand \ket [1] {|{#1}\rangle}
\newcommand \bra [1] {\langle {#1}|}
\newcommand{\secn}[1]{Section~\ref{#1}}
\def \e {\mbox{e}}

\def \fracs #1#2 {\mbox{\small $\frac{#1}{#2}$}}

\def \bin #1#2 {{\left({#1}\atop{#2}\right)}}
\def\lapproxeq{{\ \lower 0.6ex \hbox{$\buildrel<\over\sim$}\ }}
\def\gapproxeq{{\ \lower 0.6ex \hbox{$\buildrel>\over\sim$}\ }}

\def \as {\relax\ifmmode\alpha_s\else{$\alpha_s${ }}\fi}

\def \al #1 {\frac {\as({#1})}{\pi} }
\def \ds #1 {\ooalign{$\hfil/\hfil$\crcr$#1$}}
\def \MS {\overline{\rm MS}}

\def\eq#1{Eq.~(\ref{#1})}
\def\e{\epsilon}
\def\b{\beta}
\def\f{{\rm f}}
\def\M{{\cal M}}

\def\ifm{\ifmmode}

\newcommand{\JHEP}[1]{JHEP {\bf #1}}
\newcommand{\NP}[1]{Nucl.\ Phys.\ {\bf #1}}
\newcommand{\PL}[1]{Phys.\ Lett.\ {\bf #1}}

\newcommand{\PR}[1]{Phys.\ Rev.\ {\bf #1}}

\newcommand{\PRL}[1]{Phys.\ Rev.\ Lett.\ {\bf #1}}

\title{Universal structure of subleading infrared poles in gauge theory
amplitudes}

\author{Lance J. Dixon\\
Stanford Linear Accelerator Center, Stanford University,\\
Stanford, CA 94309, USA \\
 E-mail: \email{lance@slac.stanford.edu}}
\author{Lorenzo Magnea\\
Dipartimento di Fisica Teorica, Universit\`{a} di Torino, and\\
INFN, Sezione di Torino, Via P. Giuria 1, I-10125 Torino, Italy\\
 E-mail: \email{magnea@to.infn.it}}
\author{George Sterman\\
C. N. Yang Institute for Theoretical Physics, Stony Brook University, \\
SUNY, Stony Brook, NY 11794-3840, USA \\
 E-mail: \email{sterman@max2.physics.sunysb.edu}}

\abstract{We study the origin of subleading soft and collinear poles
of form factors and amplitudes in dimensionally-regulated massless
gauge theories. In the case of form factors of fundamental fields,
these poles originate from a single function of the coupling, denoted 
$G(\as)$, depending on both the spin and gauge quantum numbers 
of the field. We relate $G(\as)$ to gauge-theory matrix elements 
involving the gluon field strength. We then show that $G(\as)$ is the 
sum of three terms: a universal eikonal anomalous dimension, a 
universal non-eikonal contribution, given by the coefficient 
$B_\delta (\as)$ of  $\delta(1 - z)$ in the collinear evolution kernel, 
and a process-dependent short-distance coefficient function, which 
does not contribute to infrared poles. Using general results on the 
factorization of soft and collinear singularities in fixed-angle massless 
gauge theory amplitudes, we conclude that all such singularities are 
captured by the eikonal approximation, supplemented only by the 
knowledge of $B_\delta (\as)$. We explore the consequences of our 
results for conformal gauge theories, where in particular we find a simple 
exact relation between the form factor and the cusp anomalous 
dimension.}


\preprint{YITP-SB-08-16 \\ DFTT-06/2008 \\ SLAC--PUB--13141}

\begin{document}

\section{Introduction}
\label{intro}

The structure of soft and collinear singularities in perturbative QCD has 
been studied in depth for decades, uncovering a pattern of exponentiation 
dictated by gauge invariance and factorization. The prototype amplitude 
for these studies is the electromagnetic form factor of a colored particle, 
whose simple color structure and renormalization properties make it an 
ideal laboratory for isolating and evaluating long-distance contributions.
In fact, the color-singlet QCD form factors of quarks and gluons can both 
be expressed in an elegant exponentiated form, each in terms of only two 
functions of the running coupling. In this paper, we will extract and interpret 
the universal functions that control subleading soft and collinear poles for 
the form factors. The same functions, as we will see, also control collinear 
poles for the full class of dimensionally-regulated fixed-angle scattering 
amplitudes.

Color-singlet parton form factors are the simplest amplitudes exhibiting 
the double logarithmic (`Sudakov') behavior characteristic of gauge 
theories in the massless limit. Following the early studies in the abelian 
theory~\cite{suda}, which were performed at leading logarithmic (LL) 
accuracy, the form factors of non-abelian gauge theory were shown to
exponentiate to arbitrary logarithmic 
accuracy~\cite{Mueller:1979ih,Collins:1980ih,Collins:1981uk,Sen:1981sd}. 
Exponentiation occurs because the form factor obeys an evolution 
equation, which in turn is a consequence of factorization and gauge 
invariance \cite{Contopanagos:1996nh}. 
Solving the evolution equation yields an especially transparent 
answer~\cite{Magnea:1990zb} if one employs dimensional regularization
as an infrared regulator, as is routinely done in finite-order perturbative 
calculations. 

Dimensional regularization, in this context, displays several remarkable 
features, going well beyond its properties of preserving gauge symmetry 
and simplifying calculations in massless theories. When performing a
resummation, in fact, dimensional regularization expresses the solution to 
the appropriate evolution equation in terms of the $d$-dimensional running 
coupling, which vanishes in the infrared for $d>4$ as a consequence of 
dimensional counting. On the one hand, this allows one to solve the 
equation in terms of a simple initial condition, since all perturbative 
contributions to the amplitude vanish as a power of the hard scale for 
$d>4$; as a consequence, the resummed amplitude can be directly 
compared to finite-order Feynman diagram calculations. On the other hand, 
the $d$-dimensional running coupling in general displays a Landau pole with 
a nonvanishing imaginary part, which in turn allows an explicit evaluation of 
the resummed amplitude in terms of analytic functions of the coupling 
$\alpha_s$ and the dimension $d$~\cite{Magnea:2000ss}.

Fixed-angle scattering amplitudes also have poles in dimensional
regularization, and more generally double-logarithmic infrared
enhancements at high energy.    Evolution
equations for on-shell high-energy scattering
were developed first for theories without gauge bosons
\cite{nongauge}, while the infrared divergences associated
with external colored particles remained an obstacle
\cite{Callan:1974zy}. The recognition that soft and 
collinear (virtual) radiation can be factorized 
in a universal way from a hard QCD process (and specifically the 
result that soft radiation can be factorized from harder collinear 
radiation~\cite{Collins:1989gx}) made it possible to generalize
the exponentiation of the form factor to amplitudes with multiple colored 
legs~\cite{Sen:1982bt,Sterman:2002qn,MertAybat:2006mz}.  

Any gauge theory scattering amplitude can be treated as a vector 
in the space of available color configurations~\cite{catani96},
\begin{eqnarray}
\label{amp}
  \M^{[\f]}_{\{r_i\}}\left(\beta_j,\frac{Q^2}{\mu^2},\as(\mu^2),
  \epsilon \right) & = &
  \sum_{L = 1}^{N^{[\f]}} \M^{[\f]}_{L}
  \left(\beta_j,\frac{Q^2}{\mu^2},\as(\mu^2),\epsilon \right)
  \, \left(c_L\right)_{\{r_i\}}
  \nonumber \\
  &=& \left| {\cal M}^{[\f]} \right \rangle\, .
\end{eqnarray}
Here the vector $\left| {\cal M}^{[\f]} \right \rangle$, 
associated with a scattering process with flavor structure labeled by 
$[\f]$, is represented by the coefficients $\M^{[\f]}_L$, in a
basis defined by a set of $N^{[\f]}$ color tensors $c_L$; $\beta_j$ 
are particle momenta $p_j$ rescaled by a hard scale $Q$, 
for example as $p_j =  (Q/\sqrt{2}) \beta_j$.
When the amplitude describes fixed-angle scattering, which 
we represent as
\beqa
  {\f}: p_1 + p_2 \rightarrow p_3 + \dots + p_{n+2}\, ,
\label{momenta}
\eeqa
each color component $\M^{[\f]}_L$ can be factorized into a product 
of `jet' functions, describing the virtual color-singlet evolution of each 
external hard particle due to collinear radiation, times a `soft' function 
organizing the effects of long-wavelength 
radiation~\cite{Sen:1982bt,Sterman:2002qn,MertAybat:2006mz},
\begin{eqnarray}
\label{facamp}
  \M^{[\f]}_{L} \left(\beta_j, \frac{Q^2}{\mu^2}, \as(\mu^2),
  \epsilon \right) & = & \prod_{i = 1}^{n + 2}J^{[i]}
  \left(\frac{Q'{}^2}{\mu^2},\as(\mu^2),\epsilon \right)
  S^{[\f]}_{LI} \left(\beta_j, \frac{Q'{}^2}{\mu^2}, \frac{Q'{}^2}{Q^2},
  \as(\mu^2), \epsilon \right) \nonumber \\
  &\ & \hspace{0.2mm} \times \ H^{[\f]}_{I} \left(\beta_j,
  \frac{Q^2}{\mu^2}, \frac{Q'{}^2}{Q^2}, \as(\mu^2) \right) \, ,
\end{eqnarray}
where $Q'$ plays the role of a factorization scale separating soft and 
collinear momenta. The soft function, $S^{[\f]}_{LI}$, is a matrix in the
vector space spanned by the color tensors $c_L$.  It acts on a vector of 
finite coefficient functions describing the effects of highly virtual particles. 
In this context, the form factor continues to play an important role: one 
can always, in fact, choose the factorization scheme so that the `jet' 
functions for a generic amplitude may be identified with the square roots 
of the form factors of the corresponding hard partons,
\begin{eqnarray}
\label{jsudakov}
  J^{[i]} \left(\frac{Q'{}^2}{\mu^2}, \as(\mu^2), \epsilon\right)
  &=& \left[\Gamma^{[i]}
  \left(\frac{Q'{}^2}{\mu^2}, \as(\mu^2), \epsilon\right)
  \right]^{\frac{1}{2}} \, ,
\end{eqnarray}
where $\Gamma^{[i]} (Q^2)$ represents the color-singlet form 
factor\footnote{In Refs.~\cite{Sterman:2002qn,MertAybat:2006mz} 
the functions $\Gamma^{[i]}$ were denoted as $\M^{[i  \to i]}$, as in the 
notation for amplitudes above.} for parton $i$.   

The soft matrix $S^{[\f]}_{LI}$ in \eq{facamp} is responsible only 
for purely infrared single poles.  For this reason, it can be computed
using the eikonal approximation, in which hard partonic lines are
replaced by Wilson lines.  This fact leads to the important conclusion
that the non-eikonal, hard collinear singularities in an arbitrary QCD
amplitude, for fixed-angle scattering, can be completely organized 
in terms of the color-singlet form factors of quarks and gluons.
Our discussion below will focus on the non-leading pole structure 
of the form factors. 

Most of the results described above were originally derived with 
phenomenological applications in mind: long-distance singularities in the 
amplitudes, in fact, are the source of logarithmic enhancements in infrared- 
and collinear-safe cross sections near kinematic boundaries, which can have 
a sizable impact on perturbative predictions and often need to be resummed 
to all orders. Furthermore, the universal structure of long-distance singularities 
at fixed order~\cite{Catani:1998bh} provides an important test for perturbative
calculations, and is an essential ingredient in the construction of subtraction
schemes, which are necessary to compute finite jet cross sections and 
event-shape distributions. 
Non-leading logarithmic enhancements were studied in deep-inelastic 
scattering and vector boson production in
\cite{Laenen:2005uz,Idilbi:2006dg,Becher:2006mr,Becher:2007ty}.
These papers uncovered the same pattern for subleading enhancements 
that we will identify for poles in fixed-angle scattering.   We will return to 
these studies in Sec.\ \ref{ecpfff}.

Beyond immediate phenomenological consequences, our analysis has
a wide range of applicability, although derived within the context of QCD.   
It is based on universal properties of quantum field theories and of gauge 
theories in particular, and encodes general information about 
their long-distance behavior.  
In recent years, in fact, QCD results on the infrared structure of amplitudes 
have been applied to supersymmetric theories, and in particular to the 
maximally supersymmetric ${\cal N} = 4$ super-Yang-Mills theory, which 
is of great theoretical interest because of its connections with string theory
through the anti-de-Sitter-space-conformal-field-theory (AdS-CFT)
correspondence~\cite{Maldacena:1997re}.

This correspondence states that the strong-coupling, planar (large $N_c$)
limit of ${\cal N} = 4$ super-Yang-Mills theory admits a simple description
in terms of solutions of a classical gravitational theory, or of strings moving 
in a weakly curved background. On the other hand, the quantum conformal
invariance of the theory, which has a vanishing $\beta$ function to all 
perturbative orders, implies a drastic simplification in the all-order perturbative
resummation of infrared and collinear singularities. Together, these two
observations suggest that perturbation theory for planar ${\cal N} = 4$ 
super-Yang-Mills amplitudes should have remarkable properties. 

Indeed, motivated by a surprising iterative property of the two-loop 
four-gluon scattering amplitude~\cite{Bern:1997nh,Anastasiou:2003kj},
an exponential form for the all-loop scattering amplitude in planar 
${\cal N} = 4$ super-Yang-Mills theory was proposed~\cite{Bern:2005iz}.  
This conjecture was constructed to be consistent with
the all-order structure of soft and collinear divergences 
derived in Refs.~\cite{Magnea:1990zb} and~\cite{Sterman:2002qn}.
At the level of finite terms, it was tested successfully
for the three-loop four-gluon amplitude~\cite{Bern:2005iz} and two-loop
five-gluon amplitude~\cite{Cachazo:2006tj,Bern:2006vw}.
For four gluons, it received striking confirmation from the work
of Alday and Maldacena~\cite{Alday:2007hr}, who computed the 
strong-coupling planar limit directly from the AdS-CFT correspondence,
in dimensional regularization, finding the same exponential form in this limit.
The conjecture is expected to hold for five gluons; recently, however, it
has been found to break down for six gluons~\cite{BDKRSVV}. Such a 
breakdown was anticipated at strong coupling by an observed inconsistency
in a particular kinematic limit with a very large number of  
gluons~\cite{Alday:2007he}, and at two loops by an analysis of the 
high-energy limit of the six-gluon amplitude~\cite{Bartels:2008ce}.

Irrespective of the form of the finite terms, one can also extract from 
the Alday-Maldacena solution the leading strong-coupling behavior of 
the same quantities that govern infrared evolution in perturbation 
theory~\cite{Alday:2007hr,Alday:2007mf}. The leading (double) poles 
in the exponent for the amplitude, for example, are governed by the 
cusp anomalous dimension~\cite{Korchemsky:1985xj,%
Ivanov:1985np,Korchemsky:1987wg,Korchemsky:1988hd}. 
The AdS-CFT correspondence can then be tested at the level of elementary 
fields, by looking for consistency between the perturbative 
series at weak coupling and the strong-coupling limit.   

In the case of leading poles, the comparison can be made very 
precise because the cusp anomalous dimension has a well-defined 
non-perturbative definition in field theory, and a wealth of information 
is available about its properties.  For planar ${\cal N} = 4$ 
super-Yang-Mills theory, it has been computed perturbatively 
up to four loops~\cite{Bern:2005iz,Korchemsky:1987wg,%
Bern:2006ew,Cachazo:2006az,Korchemsky:1992xv}. 
A strong-coupling expansion can be derived from string theory, in 
which the first three terms have now been 
computed~\cite{Gubser:2002tv,Frolov:2002av,RTT}.
Most remarkably, one can exploit the observed integrability properties 
of the theory to construct an integral equation~\cite{Beisert:2006ez}, 
whose solution extends over all values of the coupling, and is in 
precise agreement with all four weak-coupling
and all three strong-coupling~\cite{CuspStrong} coefficients.

Subleading poles for scattering amplitudes are less well understood.  
In general, they cannot be expressed completely in terms of eikonal
amplitudes, although we will see that eikonal amplitudes still play an
important role in characterizing them.  It is worth remarking that
there has been considerable renewed interest of late in the 
perturbative study of eikonal amplitudes, or Wilson loop expectation
values, inspired by their role in the strong-coupling approach of Alday 
and Maldacena.  A close, but still not fully explained, relation between 
Wilson loop expectation values and the maximally-helicity-violating 
scattering amplitudes in ${\cal N}=4$ super-Yang-Mills theory has been
uncovered at one loop~\cite{Neq4WL1} and verified at two
loops~\cite{BDKRSVV,Neq4WL2,Neq4WL26pt}.

It is the purpose of our paper to provide a more precise characterization 
of the subleading-pole singularities in a massless gauge theory such as QCD 
or ${\cal N} = 4$ super-Yang-Mills theory. We begin by reviewing briefly 
in \secn{fofac} the known results about the Sudakov form factor, which 
we present in a simplified form and then apply to the case of a 
conformal theory.  In \secn{fafac}, we revisit the standard 
factorization~\cite{Collins:1989bt} in the context of dimensional 
regularization.  This allows us to give explicit operator expressions 
for the functions that control the poles of the form factors, and thus 
the collinear poles of all other fixed-angle amplitudes, via \eq{facamp} 
above.  In \secn{evolv}, we derive evolution equations for 
these operators, and identify their anomalous dimensions. This leads to
an explicit expression, in terms of these anomalous dimensions, for the 
function $G(\as)$, which determines the subleading poles of the form factor.
In \secn{pdfreln}, we construct another expression for $G(\alpha_s)$, 
by relating collinear singularities of the form factor, for a given parton 
species, to the virtual contributions to the corresponding 
parton distribution. This leads us to identify explicitly the only 
non-eikonal long-distance contribution to $G(\as)$, which is given by the 
virtual term of the diagonal Altarelli-Parisi splitting function for the chosen
parton, a fact that was pointed out at finite perturbative order in
Refs.~\cite{Ravindran:2004mb,Moch:2005tm}. Eikonal contributions 
to $G(\as)$, on the other hand, are related to the function responsible 
for soft single logarithms in threshold resummation for the Drell-Yan 
process~\cite{Belitsky:1998tc}.  As shown in \secn{pdfreln} and in the
Appendix, certain additional short-distance contributions to $G(\alpha_s$) 
are given entirely by running-coupling effects.  These contributions do not
give rise to poles in scattering amplitudes, and are proportional to $\e$
in a conformal gauge theory such as ${\cal N} = 4$ super-Yang-Mills theory.

We hope that our results will be helpful both in perturbative QCD studies, 
where a detailed knowledge of long-distance singularities to all orders is of 
direct phenomenological relevance, and in order to further our understanding 
of superymmetric gauge theories, where the striking discoveries of these 
recent years are beginning to map a precise connection between 
perturbation theory and strong coupling, possibly on the way to exact 
results.

\section{Gauge Theory Form Factors}
\label{fofac}

Let us begin by reviewing briefly the known results concerning the
color-singlet form factors of massless colored particles. For a quark, 
one can define the form factor as a matrix element of the electromagnetic 
current. In the time-like case, for example, one can write
\beq
  \Gamma_\mu (p_1, p_2; \mu^2, \e)  \equiv  \bra{0}
  J_\mu (0) \ket{{p_1,p_2}} = 
  \overline{v}(p_2) \gamma_\mu
  u (p_1) ~\Gamma \left( \frac{Q^2}{\mu^2}, \as(\mu^2), \e \right)~.
\label{def}
\eeq
Gluons do not couple directly to the electromagnetic current $J_\mu$, 
but their form factor can be defined analogously as a matrix element 
of a gauge-invariant operator.  A typical and useful example is the coupling 
of gluons to colorless scalar particles (such as the Higgs boson) through
an effective vertex constructed by integrating out a heavy fermion loop 
(the top quark in the standard model). In this case the form factor is defined 
through the coupling to the operator $- {\cal C}_H \Tr [G_{\mu \nu} 
G^{\mu \nu}]/2$, where $G_{\mu \nu}$ is the Yang-Mills field strength 
and ${\cal C}_H$ is a matching coefficient containing the dependence on
the mass of the heavy fermion. Note that in this case the effective operator
couples directly to soft gluons, but this does not change the structure of 
infrared singularities at leading power. The reasoning below therefore applies 
equally well to quarks and to gluons.

As was shown in
Refs.~\cite{Mueller:1979ih,Collins:1980ih,Collins:1981uk,Sen:1981sd}, and
reviewed in~Ref.~\cite{Collins:1989bt}, the momentum dependence of the 
form factor is determined by a simple evolution equation. The equation is a
consequence of the factorization of soft and collinear modes from 
highly virtual exchanged particles and from each 
other~\cite{Contopanagos:1996nh}, which in turn arises from the loss of 
quantum-mechanical coherence for processes occurring at widely separated 
scales.  It can be proven by making use of Ward identities. In 
dimensional regularization, with $d = 4 - 2 \e$ and $\e < 0$ in order to 
regulate mass divergences in the renormalized theory, the evolution equation 
takes the form
\beq
  Q^2 \frac{\partial}{\partial Q^2} \log \left[\Gamma \left( \frac{Q^2}{\mu^2}, 
  \as(\mu^2), \e \right) \right] = \frac{1}{2} \left[ K \left(\e, \as(\mu^2) 
  \right) + G \left(\frac{Q^2}{\mu^2}, \as(\mu^2), \e \right) \right]~,
\label{eveq}
\eeq
where the function $K(\e,\alpha_s)$ is a pure counterterm, while the 
function $G(\xi^2,\alpha_s,\e)$, which carries the momentum dependence, is
finite as $\e \to 0$. Furthermore, renormalization-group (RG) invariance of the form factor implies that
\beq
  \left( \mu \frac{\partial}{\partial \mu} + \beta(\e, \as) 
  \frac{\partial}{\partial \as} \right) G \left(\frac{Q^2}{\mu^2}, \as, 
  \e \right) = - \left( \mu \frac{\partial}{\partial \mu} + 
  \beta(\e, \as) \frac{\partial}{\partial \as} \right) 
  K \left( \e, \as \right) \equiv \gamma_K (\as) 
\label{gamk} ~,
\eeq
which is how the cusp anomalous dimension $\gamma_K (\as)$
comes into play in this context.

In order to solve \eq{eveq}, one needs to introduce the $d$-dimensional running
coupling, which solves the RG equation
\beq
  \mu \frac{\partial \as}{\partial \mu} = \b(\e, \as) = - 2 \e \as +
  \hat{\b} (\as)~, 
\label{dbet}
\eeq
where $\hat{\b} (\as)$ is the usual four-dimensional $\beta$ function,
\beq
  \hat{\b} (\as) = - \frac{\as^2}{2
  \pi} \sum_{n = 0}^\infty b_n \left( \frac{\as}{\pi} \right)^n~,
\label{beta}
\eeq
with $b_0 = (11C_A - 2 n_f)/3$ in our normalization. At the one-loop
level, the solution to \eq{dbet} is,
\beq
  \overline{\alpha}\left(\frac{\mu^2}{\mu_0^2},\as(\mu_0^2),\e\right) =
  \as(\mu_0^2) \left[\left(\frac{\mu^2}{\mu_0^2}\right)^\e - \frac{1}{\e}
  \left(1 - \left(\frac{\mu^2}{\mu_0^2}\right)^\e \right) \frac{b_0}{4 \pi}
  \as(\mu_0^2)\right]^{-1}~,
\label{loalpha}
\eeq 
which clearly reduces to the usual four-dimensional result as $\e \to 0$. 
Since the solution is RG invariant, we may use for the running coupling 
the simplified notation $\overline{\alpha} (\mu^2, \epsilon)$ whenever 
we do not need to adopt a specific boundary condition. 

At tree level (and to all orders in a conformal theory) $\overline{\alpha}$
scales as a power of $\mu$, $\overline{\alpha} (\mu^2, \e) \sim \mu^{- 2 \e} 
\as$. The running coupling thus vanishes in the infrared, as expected above the
critical dimension $d = 4$. Given the RG invariance of the form factor, this 
has the important consequence of providing us with a simple initial
condition for \eq{eveq},
\beq
  \Gamma \left( 0, \as(\mu^2), \e \right) = 
  \Gamma \left( 1, \overline{\alpha}\left(0, \e \right), \e \right) = 1~.  
\label{init}
\eeq
It is now straightforward to integrate \eq{eveq}, obtaining
\beqa
  & & \hspace{-1cm} \Gamma \left( \frac{Q^2}{\mu^2}, \as(\mu^2), 
  \e \right) ~=~ \exp \left\{ \frac{1}{2} \int_0^{- Q^2} 
  \frac{d \xi^2}{\xi^2} \Biggl[
  K \left(\e, \as(\mu^2) \right) \right. 
  \label{sol} \\ 
  & + & \left. G \left(-1, \overline{\alpha} 
  \left(\frac{\xi^2}{\mu^2},\as(\mu^2),\e \right), \e \right) 
  + \frac{1}{2} \int_{\xi^2}^{\mu^2} 
  \frac{d \lambda^2}{\lambda^2} \gamma_K \left(\overline{\alpha} 
  \left(\frac{\lambda^2}{\mu^2},\as(\mu^2), \e \right) \right) \Biggr] 
  \right\}~, \nonumber
\eeqa
where we integrated along the negative real axis to emphasize that the 
function $G$ is real for negative $Q^2$. 

The term proportional to $K(\e,\alpha_s)$ in \eq{sol} has an
apparent unregulated singularity due to the integration down to $\xi^2 = 0$: 
in fact, this term cancels exactly the $\xi^2$-independent terms arising
from the integration of $\gamma_K (\overline{\alpha})$. This can be shown 
to all orders because the function $K(\e,\alpha_s)$ is completely determined,
through \eq{gamk}, by the coefficients of the perturbative expansions 
of $\gamma_K$ and of the $\beta$ function~\cite{Magnea:2000ss}. 
As a consequence, all poles in \eq{sol} arise from integrations over the 
scale of the running coupling in the infrared region. This cancellation 
can be made explicit by considering the RG equation 
for the counterterm function $K(\e, \as)$, \eq{gamk}. Since $K$ has 
no explicit scale dependence, one can write 
\beq
  \mu \frac{d}{d \mu} K(\e, \as) = \b (\e, \as) \frac{\partial}{\partial \as}
  K(\e, \as) = - \gamma_K (\as)~.
\label{evoK}
\eeq
Using, once again, the vanishing of the running coupling in the infrared, one
has the boundary condition $K (\mu = 0) = 0$, so that \eq{evoK}
integrates to
\beq
  K \left(\e, \as (\mu^2) \right) = - \frac{1}{2} \int_0^{\mu^2}
  \frac{d \lambda^2}{\lambda^2} \gamma_K \left(\bar{\alpha} 
  (\lambda^2, \e) \right)~.
\label{solevoK}
\eeq
We now have two terms in \eq{sol} involving double scale integrals of the cusp
anomalous dimension. They both diverge, but using~\eq{solevoK}, exchanging
orders of integration, and choosing $\mu=Q$ they can be re-expressed as
\beq
  \hspace{-3mm} \Gamma \left( Q^2, \e \right) = \exp \left\{ \frac{1}{2} 
  \int_0^{- Q^2} \frac{d \xi^2}{\xi^2} \left[
  G \Big(-1, \overline{\alpha}  \left(\xi^2, \e \right), \e \Big) - 
  \frac{1}{2} \, \gamma_K \Big( \overline{\alpha}
  \left(\xi^2, \e \right) \Big) \, \log \left(\frac{- Q^2}{\xi^2} \right) \right] 
  \right\} \, .
\label{sol2} 
\eeq
In QCD, \eq{sol2} has phenomenological as well as theoretical interest. In 
fact, as described in Ref.~\cite{Sterman:2002qn}, it is one of the building
blocks for the analysis of mass singularities in general multiparton
amplitudes at finite perturbative orders, which in turn is of 
relevance for the calculation of infrared-safe observables at
high-energy colliders. We observe at this point that the function $G$ 
in~\eq{sol2} not only generates next-to-leading poles in $\e$ at each 
order in $\alpha_s$,  but it also serves as a complete infrared-safe 
coefficient function for the exponentiation of such poles, as well as finite 
parts, in the singlet form factor.

In the present context, it is interesting to notice that \eq{sol} drastically 
simplifies in a conformally-invariant theory such as ${\cal N} = 4$ 
super-Yang-Mills theory. We emphasize again that the evolution equation,
\eq{eveq}, is a consequence of gauge invariance and factorization, and 
hence holds for supersymmetric extensions of QCD, with, of course, different 
but related functions $K$ and $G$. For a conformal theory, $\hat{\beta} 
(\as) = 0$; the coupling then runs according to its mass dimension 
in $d = 4 - 2 \e$, so that $\bar\alpha_s (\lambda^2) \lambda^{2 \epsilon}
= \bar\alpha_s (\mu^2) \mu^{2 \epsilon}$; as a consequence, all integrals 
in \eq{sol} can be performed trivially in this case~\cite{Bern:2005iz}. 
Expanding the anomalous dimensions as
\beq
  \gamma_K (\overline{\alpha}) = \sum_{n = 1}^\infty \left( 
  \frac{\overline{\alpha}}{\pi} \right)^n \gamma_K^{(n)}~, \qquad
  G ( - 1, \overline{\alpha}, \e) = \sum_{n = 1}^\infty \left( 
  \frac{\overline{\alpha}}{\pi} \right)^n G^{(n)} (\e)~,
\label{expan}
\eeq
one finds several remarkably simple results. First of all, the counterterm
$K(\e, \as)$  has only simple poles, and is easily expressed in terms
of the perturbative coefficients of $\gamma_K$, as
\beq
  K (\e, \as) = \sum_{n = 1}^\infty \left( \frac{\as}{\pi} \right)^n 
  \frac{\gamma_K^{(n)}}{2 n \e}~.
\label{expanK}
\eeq
Next, one observes that the logarithm of the form factor has only
double and single poles in $\e$ to any order in perturbation theory,
in contrast to the situation in QCD, where the running of the 
coupling generates poles up to $\e^{- p - 1}$ at order $\as^p$. 
In fact, one finds explicitly~\cite{Bern:2005iz},
\beqa
  \log \left[ \Gamma \left( \frac{Q^2}{\mu^2}, \as(\mu^2),  \e \right) \right]
  & = & - \frac{1}{2} \sum_{n = 1}^\infty \left( \frac{\as (\mu^2) }{\pi} 
  \right)^n  \left( \frac{\mu^2}{- Q^2} \right)^{n \e} \left[ 
  \frac{\gamma_K^{(n)}}{2 n^2 \e^2} + \frac{G^{(n)} (\e)}{n \e} \right]
  \nonumber \\ & = & 
  - \frac{1}{2} \sum_{n = 1}^\infty \left( \frac{\as (Q^2) }{\pi} 
  \right)^n  {\rm e}^{- {\rm i} \pi n \e} \left[ 
  \frac{\gamma_K^{(n)}}{2 n^2 \e^2} + \frac{G^{(n)} (\e)}{n \e} \right]~,
\label{confgam}
\eeqa
displaying, as expected, exact RG invariance.

As a final remark, it is interesting to construct, in a
conformal theory, the ratio of the time-like to the space-like
form factor. This ratio was studied for QCD in Ref.~\cite{Magnea:1990zb}.
In that case it is of phenomenological relevance, since it enters the 
resummed expression  for the Drell-Yan cross section in the DIS 
factorization scheme~\cite{Parisi:1979xd,Sterman:1987aj,Eynck:2003fn}.
In the conformal case the analytic continuation can be performed 
explicitly, and one finds
\beq
  \log \left[ \frac{\Gamma(Q^2)}{\Gamma(- Q^2)} \right] = 
  \frac{{\rm i}}{2} \pi \bigg[ K \left(\e, \as (Q^2) \right)  +  
  G \left( -1, \as (Q^2), 0 \right) \bigg]
  + \frac{\pi^2}{8} \gamma_K \left( \as (Q^2) \right) + {\cal O} (\e)~.
\label{ratf}
\eeq
As observed in Ref.~\cite{Magnea:1990zb} in QCD, all poles in the 
ratio are given by an infinite phase, which in this case is simply
related to $\gamma_K$ via \eq{expanK}; the modulus squared of the 
ratio is thus finite in any gauge theory.  For a conformally-invariant gauge 
theory in $d = 4$, one finds the very simple expression
\beq
  \left| \frac{\Gamma(Q^2)}{\Gamma(- Q^2)} \right|^2 = \exp \left[ 
  \frac{\pi^2}{4} \, \gamma_K \left( \alpha_s (Q^2) \right) \right]~.
\label{exact}
\eeq
Since all quantities in \eq{exact} have a precise nonperturbative definition, 
and since it provides a finite, unambiguous resummation of perturbation theory, 
\eq{exact} can be argued to be an exact result.  It is easy to see that it agrees
with the result for the four-loop form factor ratio for quarks in 
QCD~\cite{Moch:2005id}, after setting all $\beta$ function coefficients 
to zero.

\section{Jet and Soft Functions in the Factorized Amplitude}
\label{fafac}

The general arguments for the factorization of QCD amplitudes and cross 
sections, isolating the contributions responsible for long-distance singular 
behavior, have been known for some time \cite{Collins:1989gx}.  In this 
section, we review the operator interpretation of this factorization,
as described originally by Collins~\cite{Collins:1989bt}. We then discuss 
the one-loop corrections to each of the relevant factors in dimensional
regularization.

At all orders, three basic physical principles apply to hard-scattering 
amplitudes, including the form factors under consideration. First, soft
gluons decouple from hard virtual partons at leading power in the hard scale, 
since each soft gluon insertion in a hard subdiagram adds to the diagram a 
new propagator far off the mass shell. Second, virtual hard partons collinear 
to an external hard parton effectively decouple from the remainder of the 
hard subdiagram, becoming insensitive to the energy and spin of fast partons
moving in different directions.  Finally, soft gluons decouple from jets,
because their long wavelength does not allow them to discriminate features
of a narrow jet other than its overall color and direction. 

To express an amplitude in factorized form unambiguously, we identify
operator expressions generating the leading contributions in each relevant
region in momentum space, and make subtractions appropriate to avoid
double counting. In the case of the form factor, the resulting 
factorization~\cite{Collins:1989bt} is depicted in Figure~\ref{factorfig} 
and can be expressed as
\beqa
  \Gamma \left( \frac{Q^2}{\mu^2}, \as(\mu^2), \e \right) & = &
  C \left( \frac{Q^2}{\mu^2}, \frac{(p_i \cdot n_i)^2}{n_i^2 \mu^2}, 
  \as (\mu^2), \e \right) \, \times \,
  {\cal S} \left( \beta_1\cdot \beta_2, \as (\mu^2), \e \right)
  \nonumber \\ && \times \,
  \prod_{i = 1}^2 \left[ 
  \frac{J \left( \frac{(p_i \cdot n_i)^2}{n_i^2 \mu^2},\as (\mu^2), \e \right)}
  {{\cal J} \left( \frac{(\beta_i \cdot n_i)^2}{n_i^2}, \as(\mu^2), \e \right)} 
  \right]~.
\label{factorize}
\eeqa
For definiteness, in \eq{factorize} we have in mind the 
time-like form factor 
for a massless quark, so that $p_1^2 = p_2^2 = 0$ and $(p_1 + p_2)^2 = 
Q^2$. We also define the quark and antiquark velocities $\beta_i$ via
$p_i^\mu = (Q/\sqrt{2})\beta_i^\mu$, so that $ \beta_1 \cdot \beta_2 
= 1$, while the vectors $n_i^\mu$ define the directions of auxiliary gauge 
links to be discussed below.
\FIGURE{
\resizebox{0.70\textwidth}{!}{\includegraphics{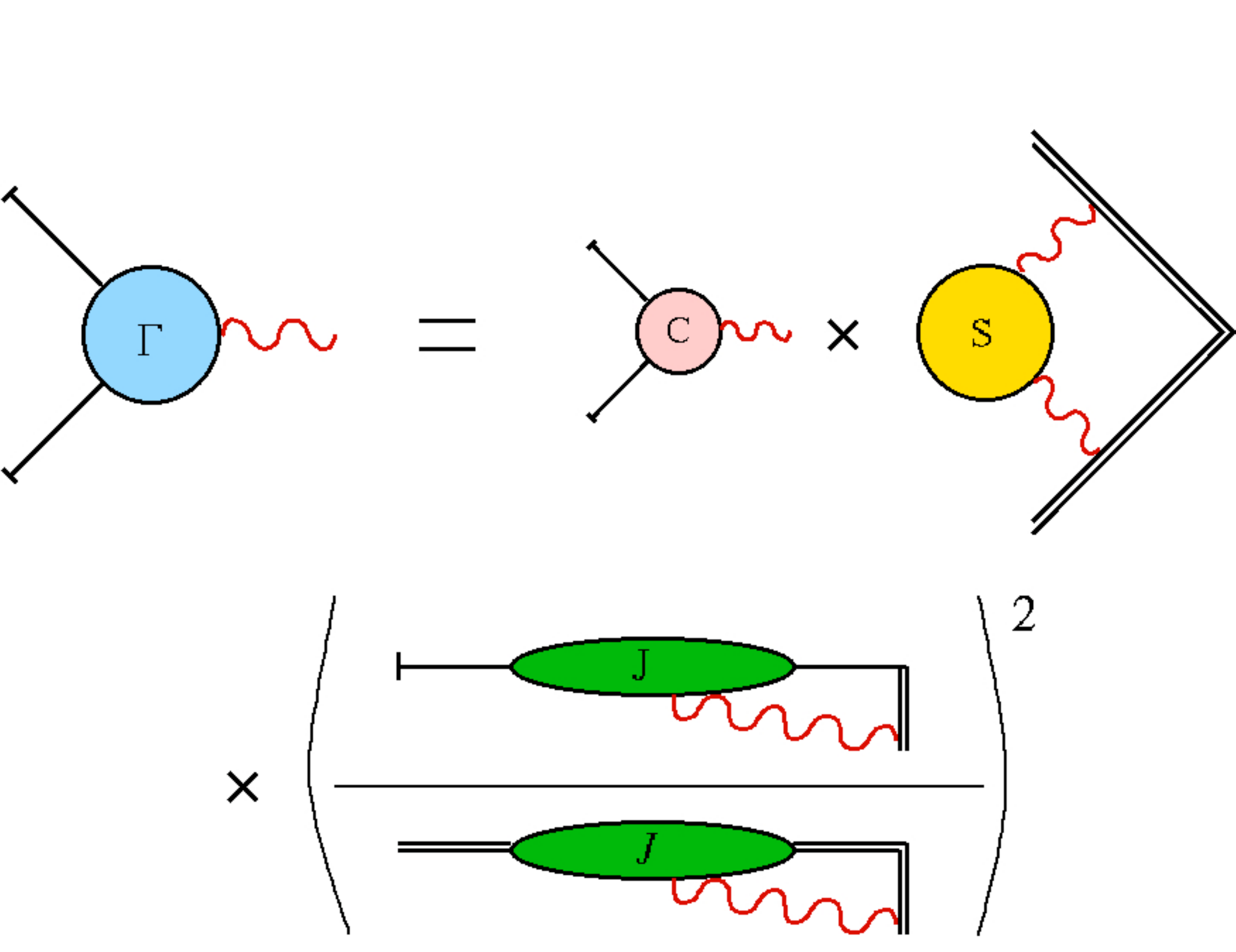}}
\caption{A graphical representation of the factorization of the form factor}
\label{factorfig} }
The hard function $C$ summarizes the short-distance contributions to the 
form factor, and is finite as $\e\to0$.  In order to define the remaining
functions appearing in \eq{factorize}, it is useful to introduce first a 
notation for the Wilson line which describes the eikonal couplings arising 
in the soft and collinear limits. These couplings are generated by the 
operator
\beq
  \Phi_n (\lambda_2, \lambda_1) =
  P \exp \left[\, {\rm i} g \int_{\lambda_1}^{\lambda_2} d \lambda \,
  n \cdot A(\lambda n) \, \right]~,
\eeq
describing a gauge link in direction $n^\mu$. 

The `partonic jets' $J$ appearing in \eq{factorize} are matrix elements 
for the transition from partonic states to the vacuum, mediated
by the corresponding partonic (in this case, quark) field.  Their operator
definition is 
\beq
  J \left( \frac{(p \cdot n)^2}{n^2 \mu^2}, \as (\mu^2), \e \right) \, 
  u (p) \, = \, \langle 0 \, | \Phi_n (\infty, 0) \, \psi (0) \,  | p \rangle\, .
\label{Jdef}
\eeq
The spinor $u(p)$ has been inserted to normalize the jet to unity at zeroth 
order;  color and Dirac indices are implicit. The gauge link in the $n^\mu$ 
direction has a double role: on the one hand, it makes the matrix element 
gauge invariant; on the other hand, it mimics the coupling of 
gluons collinear to the incoming parton (say, a quark)
to the opposite moving hard parton (say, the antiquark). 
Notice that, so long as $n^2\ne 0$,  the function $J$ (and similarly
${\cal J}$, defined below) is invariant under rescalings of the vectors 
$n_i$. Factorization makes it convenient to consider the vectors 
$n_i^\mu$ in \eq{factorize} to be space-like~\cite{Collins:1989bt};
a typical choice is $n_1 = \beta_1 - \beta_2 = - n_2$; the choice of 
vectors $n_i^\mu$ is otherwise free.   This freedom can be used to 
derive the evolution equation (\ref{eveq})~\cite{Contopanagos:1996nh}.
Clearly, $J$ has infrared divergences, as well as collinear divergences 
associated with gluons collinear to $p^\mu$.

Next, we introduce the soft function ${\cal S}$, as the vacuum
expectation value of two light-like Wilson lines in the directions $\beta_1$ 
and $\beta_2$,
\beq
  {\cal S} \left( \beta_1 \cdot \beta_2, \as (\mu^2), \e \right) =
  \langle 0 | \Phi_{\beta_2} (\infty,0) \, \Phi_{\beta_1} (0, - \infty) 
  \, | 0 \rangle~.
\label{calSdef}
\eeq
The soft function ${\cal S}$ is the eikonal limit of the full form factor, and thus
contains double poles at every order, associated with gluons that are soft and
collinear to either hard leg. It has several useful properties.
First, it is a pure counterterm in any minimal regularization scheme, 
because all its Feynman diagrams have no mass scale.
In addition, because it is defined purely in terms of Wilson lines,
it exponentiates according to the general non-abelian exponentiation
theorem~\cite{Gatheral:1983cz,Frenkel:1984pz}, and its logarithm can 
be expressed in terms of a specific subset of Feynman diagrams (`webs'), 
before the soft loop momentum is integrated over. 
An important feature of the defining matrix element (\ref{calSdef}) is
that it is invariant under boosts along the $\beta_1 \beta_2$ axis in
any frame where these velocities are back to back. On the other hand, 
while one might expect ${\cal S}$, from its operator definition,
to be invariant under rescalings of the velocities, for light-like $\beta_i$
this invariance is broken (as we will further discuss below); thus, $\cal S$ 
can depend on the velocities, through the combination 
$\beta_1 \cdot \beta_2$ only.   

There are a number of subtleties in the evaluation of purely
eikonal functions, including ${\cal S}$.   As a sum of scaleless
integrals, ${\cal S}$ vanishes before renormalization on a
diagram-by-diagram basis.   This is perhaps why classic works
on the renormalization of Wilson loops determine anomalous dimensions by
introducing explicit mass scales \cite{Korchemsky:1985xj,Ivanov:1985np,Korchemsky:1987wg}, 
such as time-like lengths $\beta_i^2>0$ and/or cutoffs on the lengths of 
the Wilson lines $\Phi_{\beta_i}$.   This method was also used
in Ref.\ \cite{MertAybat:2006mz}.
Since here we are discussing specifically dimensionally-regularized
amplitudes, we prefer to employ dimensional regularization throughout.
To do so consistently, we must identify at each order the infrared-regularized
coefficient of an ultraviolet pole.   This is possible precisely
because the logarithm of ${\cal S}$ has only a single 
ultraviolet divergence, which can be isolated systematically.

As a practical matter, to evaluate dimensionally-regulated integrals
it is convenient to rescale the light-like velocities $\beta_i$ to
have units of mass.   In this way, standard shifts of
loop momenta can be carried out.    This can be done without
loss of generality in the Wilson lines $\Phi_{\beta_i}$ that
define ${\cal S}$, by simple changes of variables.   
Below, we shall replace $\beta_i$ by $(\mu/\sqrt{2})\beta_i$,
with $\beta_1\cdot \beta_2=1$.   Then, for the  choice $\mu=Q$,
the velocities are identified with their corresponding momenta.
For now, however, we keep
the scales of the $\beta_i$ arbitrary, and will continue to refer to
them as `velocities'.  

We now turn to the exponentiation of poles in ${\cal S}$.
Collinear contributions to the soft function ${\cal S}$ 
can be factorized from purely soft contributions in the same manner as
for the full form factor, but with partonic jets replaced by eikonal jets,
for which we will provide definitions shortly. As a consequence,
${\cal S}$ satisfies an evolution equation analogous to \eq{eveq}, and 
the resulting solution is of the same form as \eq{sol} for the partonic form
factor, but with $-Q^2$ replaced by $\mu^2$ everywhere, and with an 
$\e$-independent single-log function  $G_{\rm eik}$ replacing $G$. 
The functions $\gamma_K(\as)$ and $K(\e,\as)$, on the other hand, 
are the same as in~\eq{sol}, because the eikonal form factor matches 
exactly the partonic one in all infrared-singular regions, including the 
infrared-collinear ones. The  arguments leading to \eq{sol2} still hold, and 
we are led to 
\beqa
  {\cal S} \left( \beta_1 \cdot \beta_2, \as(\mu^2), \e \right) & = & 
  \exp \left\{ \frac{1}{2} \int_0^{\mu^2} 
  \frac{d \xi^2}{\xi^2} \Bigg[ G_{\rm eik} \Big( \beta_1 \cdot \beta_2,
  \, \overline{\alpha} \left(\xi^2, \e \right) \Big) \right. \nonumber \\
  && \hspace{20mm} \left. - \frac{1}{2} \gamma_K 
  \Big(\overline{\alpha} \left(\xi^2, \e \right) \Big)  \log 
  \left( \frac{\mu^2}{\xi^2} \right) \Bigg] \right\}~.
\label{soleik}
\eeqa
Notice that the lack of explicit $\e$ dependence of $G_{\rm eik}$ 
ensures that ${\cal S}$ is a pure counterterm, consistent  with its 
diagrammatic interpretation. We expect the real part of the function 
$G_{\rm eik}$, computed in back-to-back kinematics, to be related 
to the anomalous dimension $\Gamma_{\rm DY}$, defined and computed 
at two loops in Ref.~\cite{Belitsky:1998tc}\footnote{The ``leading transcendentality'' term in $\Gamma_{\rm DY}^{(2)}$, proportional
to $\zeta(3)$,  controls the single poles of the polygonal Wilson loop 
expectation values computed recently at two loops in ${\cal N} = 4$ 
super-Yang-Mills theory~\cite{Neq4WL2}.}.

Including in the factorization both partonic jets and the full eikonal form 
factor clearly double counts the soft-collinear regions. This can be avoided 
if one divides by eikonal versions of the two jets, which are defined as
\beqa
  {\cal J} \left( \frac{(\beta_1 \cdot n_1)^2}{n_1^2}, \as(\mu^2), \e 
  \right) &=& \langle 0 | \Phi_{n_1} (\infty, 0) \, \Phi_{\beta_1}(0, - \infty)
  \, | 0 \rangle~, \nonumber\\
  {\cal J} \left( \frac{(\beta_2 \cdot n_2)^2}{n_2^2}, \as(\mu^2), \e 
  \right) &=& \langle 0 | \Phi_{\beta_2} (\infty, 0) \, \Phi_{n_2}(0, - \infty)
  \, | 0 \rangle~.
\label{calJdef}
\eeqa
These jets are also pure counterterms in dimensional regularization, 
because  they do not depend on any mass scale. They have soft-collinear enhancements from gluons moving in the $\beta_i$ directions, matching 
those of both  the soft function ${\cal S}$ or the partonic jets $J$. As a
consequence the ratio $J/{\cal J}$ has only single collinear poles at every 
order, associated with hard collinear radiation, while the ratio
\beq
  \overline{{\cal S}} \left( \rho_{12}, \as(\mu^2), \e \right)
  \equiv \frac{{\cal S} \left( \beta_1 \cdot \beta_2, \as(\mu^2), 
  \e \right)}{\prod_{i = 1}^2 {\cal J} \left( \frac{(\beta_i 
  \cdot n_i)^2}{n_i^2}, \as(\mu^2), \e \right)}
\label{Sbar}
\eeq
has only single infrared poles associated with soft gluons emitted at wide 
angles from the hard partons. In \eq{Sbar} we have noted that the function
$\overline{\cal S}$ can only depend on the homogeneous ratio
\beq 
  \rho_{12} \equiv \frac{\left(- \beta_1 \cdot \beta_2 \right)^2 \, 
  n_1^2  \, \, n_2^2}{\left( - \beta_1 \cdot n_1 \right)^2 \left( - 
  \beta_2 \cdot n_2 \right)^2} \, .
\label{rho12}
\eeq
The reason is that invariance under separate rescalings of the velocities
$\beta_i$, which was broken for ${\cal S}$, must be recovered in 
$\overline{\cal S}$, which contains the complete $\beta$ dependence 
of the form factor; furthermore, homogeneity in $n_i$ is built into the 
eikonal Feynman rules. The simplicity of these relations is a direct result 
of eikonal exponentiation. In the exponent, collinear and soft regions enter 
additively and universally~\cite{Berger:2002sv}.

The one-loop diagrams associated with the functions entering 
\eq{factorize} are easily evaluated using eikonal Feynman rules where 
appropriate.  We give the results below in the $\MS$ scheme; thus
all factors of $\log (4 \pi)$ and $\gamma_E$ are absent, having been 
absorbed into the definition of the renormalization scale $\mu$. 

The soft function receives a one-loop contribution only from 
the vertex correction diagram, since self-energies on eikonal light-like 
lines vanish like $\beta_i^2$. The full one-loop soft function is 
then given by the UV counterterm for the time-like vertex 
correction~\cite{Botts:1989kf}\footnote{Note that
the coefficient of $- \beta_1 \cdot \beta_2$ in the argument of 
the logarithm in \eq{soft1} can be changed by rescaling the eikonal 
Feynman rules. Associating with each gluon emission, for example, a factor of
$\kappa \beta_\mu/(\kappa \beta \cdot k)$, instead of the usual factor
$\beta_\mu/(\beta \cdot k)$, rescales the argument of the logarithm by a 
factor $\kappa^2$. This ambiguity is associated with the broken invariance
of the function $\cal S$ under rescalings of $\beta_i$, and corresponds 
to a choice of scheme in the renormalization of ${\cal S}$, which was 
discussed above. Once again, this ambiguity does not affect physical 
quantities: the dependence on $\kappa$ cancels between the soft 
function ${\cal S}$ and the eikonal jets ${\cal J}$, as discussed below
\eq{rho12}. Notice also that the invariance under rescalings of the vectors 
$n_i^\mu$ in the jet functions is not broken, since $n^2 \ne 0$, so that
there are no collinear divergences associated with them.},
\beq
  {\cal S}^{(1)} \left( \beta_1 \cdot \beta_2, \, \e \right) =
  - \, \frac{\alpha_s}{4 \pi} \, C_F \left[ \frac{2}{\e^2} - \frac{2}{\e} 
  \log \left( {- \beta_1 \cdot 
  \beta_2} \right) \right]~.
\label{soft1}
\eeq
Matching \eq{soleik} and \eq{soft1},  we recover, as expected, that 
$\gamma_K^{(1)} = 2 C_F$ at one loop, while\footnote{%
Equation~(\ref{geik1}) is 
consistent with Refs.~\cite{Belitsky:1998tc,Korchemsky:1993uz}, which find 
that $\Gamma^{(1)}_{\rm eik} = \Gamma^{(1)}_{\rm DY} = 0$, 
provided one chooses the subtraction scheme for collinear poles 
corresponding to $\kappa = 1$ of the footnote above. In this scheme the argument of the logarithm is $-  \beta_1 \cdot \beta_2$, 
as shown in (\ref{soft1}), so that $G^{(1)}_{\rm 
eik}$ is purely imaginary for back-to-back time-like kinematics,
giving a vanishing contribution to one-loop cross sections, while it vanishes 
for space-like kinematics. Note also that the ``DIS'' contour used in 
Ref.~\cite{Korchemsky:1993uz} differs from the space-like configuration
considered here, so $G^{(1)}_{\rm eik}$ has no correspondence
with the $\Gamma^{(1)}_{\rm DIS}$ defined there.}.
\beq
  G^{(1)}_{\rm eik} \left( \beta_1 \cdot \beta_2 \right) = - \,
  \frac{\alpha_s}{\pi} \, C_F \, 
  \log \left( - \beta_1 \cdot \beta_2 \right)~.
\label{geik1}
\eeq
The eikonal jet ${\cal J}$ receives contributions at one loop from 
both the eikonal vertex correction and the self-energy diagram 
on the eikonal line along the $n_i^\mu$ direction. 
This eikonal self energy, ${\cal J}_{n^2}$ below, is a single pole
pure  counterterm at one loop,  which is common 
to the partonic jet $J$ and to the eikonal jet ${\cal J}$.  It cancels 
in their ratio, but contributes to their respective evolution equations. 
The vertex correction to ${\cal J}$, which we denote by 
${\cal J}_{\rm V}$, also contributes only through a counterterm.
The complete one-loop eikonal jet is then
\beqa
  {\cal J}^{(1)} &=& \frac{1}{2}\, {\cal J}_{n^2}^{(1)} + 
  {\cal J}_{\rm V}^{(1)}\, , \nonumber\\
  {\cal J}^{(1)}_{\rm V} \left(\frac{(\beta \cdot n)^2}{n^2}, 
  \, \e \right) &=& - \, \frac{\alpha_s}{4 \pi} \, C_F \left[ \frac{1}{\e^2} + 
  \frac{1}{\e} \log \left( 
  \frac{n^2}{2(- \beta \cdot n)^2} \right) \right]\, ,
  \nonumber \\
  {\cal J}_{n^2}^{(1)}  &=& - \frac{\as}{2 \pi} \, C_F \, \frac{1}{\e} \, .
\label{eikj1}
\eeqa
As expected, the soft-collinear double pole of the eikonal jet is one 
half of the corresponding  pole in the eikonal form factor ${\cal S}$.
Here and below, the one-loop self-energy counterterm (for the
$n$ eikonal) is multiplied by $1/2$, which reflects the removal of 
the square root of the residue of the relevant two-point function in 
a normalized $S$-matrix element.

Turning to the partonic jet $J$, we encounter in its vertex correction 
a one-loop diagram that is not simply a pure counterterm. The full
one-loop result is the sum
\beq
  J^{(1)} =  \frac{1}{2}\, {\cal J}^{(1)}_{n^2} + J^{(1)}_{\rm V}
  + \frac{1}{2}\, J_{\rm P}^{(1)} \, , \label{allJ1}
\eeq
with ${\cal J}_{n^2}^{(1)}$ given as above, while $J_{\rm V}^{(1)}$ is 
the quark-eikonal vertex correction, given by 
\beqa
  J^{(1)}_{\rm V} \left(\frac{(p \cdot n)^2}{n^2 \mu^2}, 
  \, \e \right) & = & - \, \frac{\alpha_s}{4 \pi} \, C_F \left[ \frac{1}{\e^2} + 
  \frac{1}{\e} \left( 2 + \log \left( 
  \frac{n^2 \mu^2}{(- 2 p \cdot n)^2} \right) \right) \right.
  \nonumber \\
  && \hspace{-2cm} + \, \left. \frac{1}{2} \log^2 \left( 
  \frac{n^2 \mu^2}{(- 2 p \cdot n)^2} \right) +
  \log \left( \frac{n^2 \mu^2}{(- 2 p \cdot n)^2} 
  \right)  + 2 + \frac{5}{12} \pi^2 
  + {\cal O} \left( \e \right) \right]~;
\label{jvert1}
\eeqa
finally, $J_{\rm P}^{(1)}$ is the pure counterterm self-energy on the quark leg,
\beq
  J^{(1)}_{\rm P} \left( \e \right) =  \frac{\alpha_s}{4 \pi} \, C_F \, 
  \frac{1}{\e}~.
\label{jprop1}
\eeq
Note that, as expected, the vertex correction $J_{\rm V}^{(1)}$ gives
the same double pole as the eikonal jet, \eq{eikj1}.   

Collecting all the ingredients, and applying \eq{factorize}, we expect to 
reproduce all the infrared and collinear poles of the form factor 
at one loop. Indeed we find
\beqa
  \Gamma^{(1)}_{\rm pole} \left( \frac{Q^2}{\mu^2}, \e \right) & = & 
  {\cal S}^{(1)} \left( \beta_1 \cdot \beta_2, \, \e \right) +  
  J^{(1)}_{\rm V,\, pole} \left(\frac{(p_1 \cdot n_1)^2}{n_1^2 \, \mu^2},  
  \, \e \right) +
  J^{(1)}_{\rm V,\, pole} \left(\frac{(p_2 \cdot n_2)^2}{n_2^2 \, \mu^2},  
  \, \e \right)   \nonumber \\ & & - \, \,
  {\cal J}^{(1)}_{\rm V} \left(\frac{(\beta_1 \cdot n_1)^2}{n_1^2}, 
  \, \e \right) -
  {\cal J}^{(1)}_{\rm V} \left(\frac{(\beta_2 \cdot n_2)^2}{n_2^2}, 
  \, \e \right) +
  J^{(1)}_{\rm P} \left( \e \right) \\
  & = & \frac{\alpha_s}{4 \pi} \, C_F \left[ - \frac{2}{\e^2} - 
  \frac{2}{\e} \log \left( \frac{\mu^2}{- Q^2} \right)
  -  \frac{3}{\e} \right]~, \nonumber
\label{gamma1p}
\eeqa
which matches the pole structure of the one-loop form factor in 
dimensional regularization,
\beq
  \Gamma^{(1)} \left( \frac{Q^2}{\mu^2}, \e \right) = 
  - \, \frac{\alpha_s}{4 \pi} \, C_F \left( \frac{\mu^2{\rm 
  e}^{\gamma_E}}{- Q^2} \right)^\e
  \frac{\Gamma^2 (1 - \e) \Gamma(1 + 
  \e)}{\Gamma(1 - 2 \e)} 
  \left( \frac{2}{\e^2} + \frac{3}{\e} + 8  + {\cal O} (\e) \right) ~.
\label{gamma1}
\eeq
Equation~(\ref{gamma1p}) also implies that
\beq
  G^{(1)} \left( \frac{Q^2}{\mu^2}, \e \right) = 
  \frac{\alpha_s}{\pi} \, C_F \, \left[ \log \left(\frac{\mu^2}{- Q^2} 
  \right) + \frac{3}{2} + {\cal O} (\e) \right] \, ,
\label{G1}
\eeq
in agreement with Ref.~\cite{Magnea:1990zb}. Our task is now to construct 
an all-order expression for the function $G$ in terms of the anomalous 
dimensions of the various functions building up the form factor according 
to \eq{factorize}.

\section{From Factorization to an Operator Intepretation for $G(\alpha_s)$}

\label{evolv}

Let us begin our investigation of $G(\as)$
by considering the renormalization properties of the various 
functions entering \eq{factorize}. The partonic jet function $J$ and the
short-distance function $C$ are multiplicatively renormalizable, with 
anomalous dimensions depending on the coupling $\as$ but not
on the infrared regulator $\e$.  Eikonal functions such as ${\cal S}$ and 
${\cal J}$, on the other hand, require extra care: in general, their anomalous
dimensions need infrared regularization in their own right,  because of the 
overlap of collinear and ultraviolet divergences in the renormalization of 
any `cusp' singularity involving light-like Wilson 
lines~\cite{Korchemsky:1987wg}.

Consider first the soft eikonal function ${\cal S}$. The observation that it 
is a pure counterterm, of the general form of \eq{soleik}, leads to
\beqa
  \mu \frac{d}{d \mu} \log{{\cal S}} \left( \beta_1\cdot \beta_2, 
  \as (\mu^2), \e \right) & = & \beta (\e, \as) \, \frac{\partial}{\partial \as}
  \log{{\cal S}} \left( \beta_1\cdot \beta_2, 
  \as (\mu^2), \e \right) \nonumber \\ & \equiv&  - \, 
  \gamma_{\cal S} \left(\beta_1 \cdot \beta_2, \as (\mu^2), 
  \e \right) \, ,
\label{evoS}
\eeqa
where the singular anomalous dimension $\gamma_{{\cal S}}$ is related 
to the function $G_{\rm eik}$ and to the cusp anomalous dimension 
$\gamma_K$ by
\beqa
  \gamma_{\cal S} \left(\beta_1 \cdot \beta_2, \as (\mu^2), 
  \e \right) &=& \, - \, G_{\rm eik} \left( \beta_1 \cdot \beta_2, 
  \as (\mu^2) \right) + \frac{1}{2} \int_0^{\mu^2} \frac{d \xi^2}{\xi^2} 
  \gamma_K \left( \overline{\alpha} (\xi^2, \e) \right) \nonumber\\
  & = & - \, G_{\rm eik} \left( \beta_1 \cdot \beta_2, \as (\mu^2) 
  \right)  - K \left(\e, \as (\mu^2) \right)\, .
\label{gammaS}
\eeqa
Here, in the second equality, we have used \eq{solevoK}.
We observe that $\gamma_{\cal S}$ has a single infrared pole 
determined by the cusp anomalous dimension $\gamma_K$. In order 
to work with infrared-finite anomalous dimensions, one can instead
consider the function $\overline{\cal S}$, defined in Eq.\ (\ref{Sbar}),
which carries only single infrared poles
due to wide angle soft radiation. It obeys
\beq
  \mu \frac{d}{d \mu} \log{\overline{\cal S}} \left( \rho_{12}, 
  \as (\mu^2), \e \right) =  - \ \gamma_{\overline{\cal S}} 
  \left(\rho_{12}, \as (\mu^2) \right) \, .
\label{gammabarS}
\eeq
Finally, one may define
\beqa
  \mu \frac{d}{d \mu} 
  \log J \left( \frac{(p_i \cdot n_i)^2}{n_i^2 \mu^2},\as (\mu^2), \e \right) 
  & = & - \, \gamma_J (\as) \,, \nonumber \\
  \mu \frac{d}{d \mu} \log C \left( \frac{Q^2}{\mu^2}, 
  \frac{(p_i \cdot n_i)^2}{n_i^2 \mu^2}, \as (\mu^2), \e \right)
  & = & - \, \gamma_C(\rho_{12}, \as) \,,
\label{gammaCJ}
\eeqa
where the functional dependence of $\gamma_C$ on $\mu$ is dictated 
by the requirement that the form factor as a whole not be renormalized, 
which implies, using Eqs.\ (\ref{factorize}) and (\ref{Sbar}),
\beq
  \gamma_{\overline{\cal S}} \left( \rho_{12}, \as \right) +
  \gamma_C \left( \rho_{12}, \as \right) + 2 \gamma_J 
  \left( \as \right) = 0 \, .
\label{gammasum}
\eeq
At one loop, we can compute $\gamma_J$ by combining
the terms specified in \eq{allJ1}; similarly,  $\gamma_{\bar{\cal S}}$
is derived from the one-loop results for ${\cal S}$ and ${\cal J}$, combined
as in \eq{Sbar}.   We find
\beqa
  \gamma_{\overline{\cal S}}^{(1)} \left( \rho_{12} \right) & =  &
  \frac{\as}{\pi} \, C_F \, 
  \left[1 + \frac{1}{2} \log \left(\frac{\rho_{12}}{4}\right) \right] 
  \, , \nonumber \\
  \gamma_J^{(1)} & = & - \frac{\as}{\pi} \,
  \frac{3}{4} \, C_F \, ,
\label{gamJsbar} 
\eeqa
where we note that $\gamma_J^{(1)}$ equals the one-loop anomalous 
dimension of the quark field. We then derive for $\gamma_C^{(1)}$,
using Eq.\ (\ref{gammasum}),
\beqa
  \gamma_C^{(1)} \left( \rho_{12} \right) & =  &
  \frac{\as}{\pi} \, C_F \, \left[ \frac{1}{2} - \frac{1}{2} \log 
  \left(\frac{\rho_{12}}{4}\right) \right] \, .
\label{oneloga}
\eeqa
Having exhibited these one-loop examples, we continue with
the general discussion.

The next step is to consider the dependence on the vectors $n_i^\mu$, 
which enter the form factor through the jet functions $J$ and 
their eikonal counterparts ${\cal J}$. Following the reasoning of 
Refs.~\cite{Collins:1981uk,Sen:1981sd} in axial gauge, generalized to 
arbitrary gauges in Ref.~\cite{Contopanagos:1996nh}, we begin by observing
that the form factor must be independent of $n_i^\mu$. Defining
$x_i \equiv \left(- \beta_i \cdot n_i \right)^2/n_i^2$, we can write
\beq
  x_i \, \frac{\partial}{\partial x_i} \log \Gamma\left(\frac{Q^2}{\mu^2}, 
  \as (\mu^2), \e \right) = 0 \, .
\label{noxdep}
\eeq
When we apply this consistency condition to the factorized cross section
in \eq{factorize}, derivatives with respect to $n_i^\mu$ 
of the partonic jet functions $J_i$ separate into ultraviolet- and 
infrared-dominated terms, according to
\beqa
  x_i \, \frac{\partial}{\partial x_i} \log J_i & = &
  \ - \, x_i \, \frac{\partial}{\partial x_i} \log C  \, + \,
  x_i \, \frac{\partial}{\partial x_i} \log {\cal J}_i \nonumber \\
  & \equiv & \frac{1}{2} \left[{\cal G}_i \Big( x_i, \as(\mu^2), \e \Big)
  + {\cal K} \Big( \as(\mu^2), \e \Big)  \right] \, ,
\label{dJdn}
\eeqa
where the second line defines the sum of the functions ${\cal G}_i$ and 
${\cal K}$. As is clear from this definition, and in analogy with \eq{eveq}, 
the function ${\cal G}_i$ carries the scale dependence, but is finite as 
$\e \to 0$, while ${\cal K}$ is a pure counterterm. At one loop we find,
directly from \eq{allJ1},
\beqa
  {\cal G}_i^{(1)} \left( x_i, \e \right) &=& \frac{\as}{2 \pi} C_F
  \left( \log \frac{n_i^2\,\mu^2}{4 (p_i\cdot n_i)^2} + 1  \right)
   + {\cal O}(\epsilon) \, , \nonumber\\
  {\cal K}^{(1)} (\e) &=& \frac{\as}{2 \pi} \, C_F \, \frac{1}{\e} \, .
\label{calGi}
\eeqa
We can now relate the functions ${\cal G}_i$ and ${\cal K}$ to
matrix elements of fields in the presence of Wilson lines. 
In fact, both partonic and eikonal jets depend on 
$n^\mu$ and on the velocity $\beta^\mu$ only through the combination 
$x = (- \beta \cdot n)^2/n^2$. One can thus simply relate their $x$ 
dependence, given in \eq{dJdn}, to their $n^\mu$ dependence, using
\beq
  p \cdot n \, \frac{\partial  J}{\partial \, p \cdot n} = 
  - \frac{n^2}{p \cdot n} \, \, p^\nu  \, \frac{\partial J}{\partial \, n^\nu} \, ,
\label{cghton}
\eeq
and similarly for ${\cal J}$, with $p^\mu$ replaced by $\beta^\mu$.
From the definitions of the jet functions, Eqs.~(\ref{Jdef}) and (\ref{calJdef}),
and from the behavior of an ordered exponential under variation with 
respect to the curve, we readily find expressions for the derivatives of 
the jet functions  in \eq{dJdn}, extending results found by Collins in 
QED~\cite{Collins:1989bt}. 

For both partonic and eikonal jets, a derivative with respect to the
vector $n^\mu$ replaces an ordered exponential in the $n^\mu$ direction
by the integral of a field strength over the original path, sandwiched 
between Wilson lines.  Written explicitly, we can use this result to 
determine the $p\cdot n$ dependence of the jet functions (\ref{Jdef}) 
in terms of matrix elements, as
\beqa
  && p \cdot n \, \frac{\partial}{\partial \, p\cdot n} \, \left[
  \log \, J \left( \frac{(p \cdot n)^2}{\mu^2 n^2}, \as (\mu^2), 
  \e \right) \right] \nonumber \\
  && \hspace{10mm}
  =  \, - \, \frac{n^2}{p \cdot n} \, J^{-1} \, \frac{p^\nu}{2 p_0} \,
  u^\dagger(p) \,
  \langle 0 | \, \frac{\partial}{\partial n^\nu} \, \Phi_n (\infty, 0) \, \psi (0) 
  \, | p \rangle \nonumber \\
  && \hspace{10mm} = \, - \, \frac{n^2}{p \cdot n} \,
  J^{-1} \, \frac{1}{2 p_0} \, u^\dagger(p) \, 
  \int_0^\infty d \lambda \, \lambda \, \langle 0 |  \,
  \Phi_n (\infty, \lambda) \, \,  p^\mu  \, n^\nu \, \left( {\rm i} g \, 
  F_{\mu \nu} (\lambda n) \right) \, \Phi_n (\lambda, 0) \, \psi (0) 
  \, | p \rangle \nonumber \\
  && \hspace{10mm} \equiv  \, {\cal G} \left( 
  \frac{(p \cdot n)^2}{\mu^2 n^2}, \as (\mu^2), \e \right) + 
  {\cal K} \left( \as (\mu^2), \e \right) \, ,
\label{Jderiv}
\eeqa
where the final equality is simply a restatement of Eq.\ (\ref{dJdn}).
Similarly, as the first equality in (\ref{dJdn})
makes clear, ${\cal K}$ is found directly from ${\cal J}$ as
\beqa
  \beta \cdot n \, \frac{\partial}{\partial \, \beta \cdot n} \,
  \log \, {\cal J} \left( \frac{(\beta \cdot n)^2}{n^2}, \as(\mu^2), \e 
  \right)
  & = & \, - \, \frac{n^2}{\beta \cdot n} \, {\cal J}^{-1} \, \beta^\nu \,
  \langle 0 | \, \frac{\partial}{\partial n^\nu} \, \Phi_n (\infty, 0) \, 
  \Phi_\beta (0, - \infty) \, | 0 \rangle \nonumber \\
  && \hspace{-60mm} = \, - \, \frac{n^2}{\beta \cdot n} \,
  {\cal J}^{-1} \, \int_0^\infty d \lambda \, \lambda \, \langle 0 |  
  \Phi_n (\infty, \lambda) \, \,  \beta^\mu n^\nu \, \left( {\rm i} g \,
  F_{\mu \nu} (\lambda n) \right) \, \Phi_n (\lambda, 0) \, \Phi_\beta 
  (0, - \infty) \, | 0 \rangle \nonumber \\
  && \hspace{-60mm} \equiv \, {\cal K} \left( \as (\mu^2), \e \right) \, .
\label{calJderiv}
\eeqa
In the perturbative expansions of these matrix elements,
the field strength operator prevents an unphysical 
gluon from coupling to the $n^\mu$ eikonal.  At each order, the 
corresponding vertex cannot appear in any jet-like subdiagram that provides 
a collinear pole. It can, however, appear in subdiagrams that carry ultraviolet
and infrared momenta \cite{Collins:1981uk}, and, in the latter case,
are associated with an infrared pole in dimensional regularization. 
The inverse jet and eikonal-jet factors multiplying the 
matrix elements cancel the residual collinear singularities, which 
factorize. The remaining terms give, order-by-order, the ${\cal G}$ 
and ${\cal K}$ functions above, from the short- and long-distance 
non-collinear regions to which the field strength vertex can contribute.
The function ${\cal G}$ is seen to be the difference of two gauge-invariant 
matrix elements, both involving the field strength and Wilson lines,
derived from, and normalized by, the partonic and eikonal jet functions.

It is worth noting that the $n^2\rightarrow 0$ limits
of Eqs.\ (\ref{Jderiv}) and (\ref{calJderiv}) are singular.
In this limit, the matrix elements become boost-invariant,
and at the same time develop collinear singularities 
for gluons in the $n^\mu$ direction.  As for the
soft function ${\cal S}$, the jet functions,
which are sums of pole contributions only, are no longer
scale invariant in the light-like vectors $\beta$ and $n$.

We can now turn to the determination of the $Q$ dependence of the 
full form factor, \eq{eveq}. $\Gamma$ depends on $Q$ directly through 
the short-distance function $C$, and indirectly through the partonic jets 
$J_i$, which depend on external momenta through $p_i \cdot n_i$.
From the factorized expression, \eq{factorize}, using \eq{gammasum},
one easily derives
\beqa
  Q \frac{\partial}{\partial Q} \log \Gamma =
  \beta(\e, \as ) \, \frac{\partial}{\partial \as} \log C - 
  \gamma_{\bar{\cal S}} - 2 \, \gamma_J + 
  \sum_{i = 1}^2 \left({\cal G}_i + {\cal K} \right) \, .
\label{evolcalGK}
\eeqa
Because the pole terms ${\cal K}(\e, \as)$ are independent of the 
kinematic variables, they are equal, and we have $K = 2 \, {\cal K}$.   
Comparing \eq{evolcalGK} with the original evolution equation, 
\eq{eveq}, we finally find an expression for $G$, in terms of the
anomalous dimensions of the soft and eikonal jet functions, and in terms 
of the functions ${\cal G}_i$, defined by the matrix elements of
Eqs.~(\ref{Jderiv}) and (\ref{calJderiv}). Explicitly, we have
\beqa
  G(\as) = \beta(\e, \as ) \, \frac{\partial}{\partial \as} \log C - 
  \gamma_{\bar{\cal S}} - 2 \gamma_J +  \sum_{i = 1}^2 {\cal G}_i \, ,
\label{Gsum}
\eeqa
which is easily verified to hold at one loop, making use of the results of 
\secn{fafac}.

\section{Eikonal and Collinear Contributions to Subleading Poles}
\label{pdfreln}

The single poles of form factors are generated by functions 
$G(\alpha_s)$, one for each type of parton, which have been 
related to gauge theory matrix elements via \eq{Gsum}.  In this section, 
we will link $G(\alpha_s)$ to two anomalous dimensions, one 
stemming from the collinear evolution of parton distributions, 
the other from the eikonal form factor, ${\cal S}$.
A relation following from this structure was verified empirically to three loops
in Ref.~\cite{Moch:2005tm}, based on an earlier observation 
of Ref.~\cite{Ravindran:2004mb}.  A similar connection has been established
at finite order between single-logarithmic contributions to the Drell-Yan cross 
section and collinear evolution kernels in Ref.~\cite{Laenen:2005uz}. These
empirical observations are established here to all orders in perturbation theory,
exploiting a connection between the form factor and parton-in-parton
distributions, which follows from factorization and which was noted already
in Ref.~\cite{Laenen:2000ij}.    To derive our result, we first have a look
at the analog of the form factor in parton evolution.

\subsection{Factorization for Virtual Contributions to Parton Distributions}

Let us begin by considering the standard definition of the light-cone 
distribution for a parton of flavor $i$, carrying momentum fraction $x$,
in a parent parton of the same flavor. For a quark, for example, one writes
\beq
  \phi_{q/q} \left(x, \e \right) = 
  \frac{1}{4 N_c} \int \frac{d \lambda}{2 \pi} {\rm e}^{- {\rm i} 
  \lambda x p \cdot \beta} \bra{p} \overline{\psi}_q (\lambda \beta) \,
  \gamma \cdot \beta \, \Phi_\beta (\lambda, 0) \,
  \psi_q (0) \ket{p} \, ,
\label{painpa}
\eeq
where $p$ is the momentum of the parent quark, which we can take, say, 
along the $(+)$ direction; $\beta$ is then an auxiliary light-cone vector
along the $(-)$  direction, and the Dirac projector $\gamma \cdot 
\beta$ selects the relevant components of the quark field, while the Wilson 
line $\Phi_\beta$ ensures gauge invariance. An analogous definition applies 
for the gluon-in-gluon distribution, with the Wilson line in the adjoint 
representation. Note that there is no explicit $p\cdot \beta$ dependence 
in the parton distributions when they are defined in this boost-invariant 
fashion in the $\overline{\rm MS}$ prescription.

In order to single out the virtual contributions to the parton distribution,
we proceed as follows. The gauge link can be split by extending it to light-like
infinity along the $\beta$ direction, according to
\beq
  \Phi_\beta (\lambda, 0) = \Phi_\beta (\lambda, \infty) 
  \Phi_\beta (\infty, 0) \, ;
\label{splitlink}
\eeq
one can now insert a complete set of states between the two Wilson lines, 
and then isolate the contribution of the vacuum. This gives the 
virtual contribution to the parton distribution at the amplitude level
as the correlator
\beqa
  \overline{\Gamma}_{q/q} \left( \frac{p \cdot \beta}{\mu}, 
  \as(\mu^2), \e   \right)
  \equiv \bra{0} \Phi_\beta (\infty, 0) \, \psi_q(0)\, \ket{ p} \, ,
\label{parton_oe}
\eeqa
coupling a single-particle state to the vacuum through the action of the 
partonic field $\psi_q$ and of a gauge link in the same color representation.
We shall define this matrix element as a sum of pure pole terms,
consistent with its interpretation as part of an $\overline{\rm MS}$ 
parton distribution function.

Clearly, the amplitude $\overline{\Gamma}_{q/q}$ in \eq{parton_oe} is 
closely related to the partonic jet, $J$, in \eq{Jdef}. In fact, the only 
difference is that the gauge link is now in a light-like direction opposite 
to the parton momentum. As a consequence, $\overline{\Gamma}_{q/q}$ 
can be factorized in the same manner as the full partonic amplitude, 
\eq{factorize}, into short-distance, jet, and soft functions. Now, however, 
we need a separate partonic jet only for the incoming line, since the 
collinear singularities of the outgoing gauge link match the collinear 
singularities of the soft function ${\cal S}$ in \eq{factorize}.  
We can then write
\beqa
  \overline{\Gamma}_{q/q} \left(\frac{p \cdot \beta}{\mu}, 
  \as(\mu^2), \e \right)
  &=& {\cal S} \left( \beta_p \cdot \beta, \as (\mu^2), \e \right) \, 
  \nonumber\\ & \ & \null \times
  C_J\left( \frac{(p \cdot n)^2}{n^2 \mu^2}, \as (\mu^2), 
  \e \right)\,
  \left[ \frac{J \left( \frac{(p \cdot n)^2}{n^2 \mu^2}, \as (\mu^2), 
  \e \right)}{{\cal J} \left( \frac{(\beta_p \cdot n)^2}{n^2}, \as(\mu^2), 
  \e \right)} \right] \, ,
\label{factor_parton_oe}
\eeqa
where we have introduced the velocity four-vector $\beta_p$ associated 
with the momentum $p$.   The function $C_J$ is a short-distance coefficient
chosen to cancel all terms that are finite for $\epsilon\to 0$ in $J$, because
$\overline{\Gamma}_{q/q}$ is defined as a sum of pole terms only.
We are assured that the function $C_J$ exists, because of the exponentiation 
of all pole terms. Using the fact that ${\cal S}$ and ${\cal J}$ are also pure 
pole terms, we may write,
\beq
  C_J\left( \frac{(p \cdot n)^2}{n^2 \mu^2}, \as (\mu^2), 
  \e \right)\,  J \left( \frac{(p \cdot n)^2}{n^2 \mu^2}, \as (\mu^2), \e 
  \right)\,  = \left[ J \left( \frac{(p \cdot n)^2}{n^2 \mu^2},
  \as (\mu^2), \e \right) \, \right]_{\rm pole} \, .
\label{Jpoles}
\eeq
Because $\overline{\Gamma}_{q/q}$ is $n$-independent,
all $n$-dependence in poles on the right-hand side of \eq{factor_parton_oe}
is guaranteed to cancel in the ratio of jet functions, 
leaving only finite $n$-dependence, which is cancelled by $C_J$.
This cancellation is possible simply because collinear singularities are 
independent of $n^\mu$ \cite{Collins:1981uk}, leaving only soft 
contributions, whose (exponentiating) poles match between $J$ and 
${\cal J}$, and short-distance contributions, which are cancelled 
by $C_J$. We observe that the ratio of partonic to eikonal jets in 
\eq{factor_parton_oe} is the same as in the basic factorized form,
\eq{factorize}.

Inserting \eq{Jpoles} into \eq{factor_parton_oe}, 
$\overline{\Gamma}_{q/q}$ can be represented as,
\beq
  \overline{\Gamma}_{q/q} \left( \beta_p \cdot \beta, \as(\mu^2), 
  \e \right) =  {\cal S} \left( \beta_p \cdot \beta, \as (\mu^2), 
  \e \right) \, \frac{ \left[ J \left( \frac{(\beta_p \cdot n)^2}{n^2 },
  \as (\mu^2), \e \right) \right]_{\rm pole} }{{\cal J} 
  \left( \frac{(\beta_p \cdot n)^2}{n^2}, 
  \as(\mu^2), \e \right)} \, ,
\label{factphivirt}
\eeq
where we have chosen the magnitude of the four-velocity $\beta_p$ 
so that $p = (\mu/\sqrt{2}) \beta_p$, and the scalar product $p 
\cdot \beta = (\mu/\sqrt{2}) \, \beta_p \cdot \beta$. Thus $\mu$ 
is the only remaining scale, which appears only as the argument of 
the coupling.  As usual, \eq{factphivirt} involves exponentiating double 
poles from ${\cal S}$, which cancel when combined with the real emission
contributions to the parton distribution, leaving behind only single, collinear 
poles that define the splitting functions.

Now, following Ref.~\cite{Laenen:2000ij}, we introduce the eikonal 
counterpart of the correlator~(\ref{parton_oe}), which will represent the 
virtual contribution to an eikonal parton distribution. This is naturally 
defined as the soft function ${\cal S}$ itself,
\beq
  \overline{\Gamma}_{q/q}^{\rm eik} \left( \beta_p \cdot \beta, 
  \as(\mu^2), \e \right)\equiv {\cal S} \left( \beta_p\cdot \beta, 
  \as (\mu^2), \e \right) \, ,
\label{phieikvirt}
\eeq
which, once again, is an exponential consisting entirely of pole terms.
We can now exploit the fact that $\MS$ parton distributions can be
defined, in moment space, simply as exponentials of the integrated 
collinear anomalous dimension, as was done in Ref.~\cite{Laenen:2000ij}. 
Furthermore, the eikonal approximation is accurate for 
real final-state radiation, up to inverse powers of the Mellin variable $N$.
The ratio of the virtual contribution for the parton distribution to its 
eikonal counterpart must thus be given entirely by the virtual term of 
the corresponding diagonal splitting function, $B_\delta^{[i]} (\as)$, 
whose normalization is defined by
\beq
  P_{ii}(x) = {\gamma_K^{[i]} (\alpha_s)\over2} \biggl[ {1\over1 - x} 
  \biggr]_+ + B_\delta^{[i]} (\alpha_s) \, \delta(1 - x) + 
  {\cal O} \Bigl( (1 - x)^0 \Bigr) \,.
\label{Bidef}
\eeq
Taking the ratio of \eq{phieikvirt} and \eq{factphivirt} we then find,
\beqa
  \frac{\left[J \left( \frac{(\beta_p \cdot n)^2}{n^2},\as (\mu^2), \e 
  \right)\right]_{\rm pole}}{{\cal J} \left( \frac{(\beta_p \cdot n)^2}{n^2}, 
  \as(\mu^2), \e \right)}
   & = &
  \frac{\overline{\Gamma}_{q/q} \left( \beta_p \cdot \beta, \as(\mu^2), 
  \e \right)}{\overline{\Gamma}_{q/q}^{\rm eik} \left( \beta_p \cdot \beta, 
  \as(\mu^2),  \e \right)}
   \nonumber \\ & = &
  \exp  \left[ \frac{1}{2}  \int_0^{\mu^2} \frac{d \xi^2}{\xi^2} \,
  B_\delta^{[q]} \left( \overline{\alpha} (\xi^2, \e) \right) \right] \, ,
\label{phiratio}
\eeqa
where the second  relation\footnote{There is a factor of $1/2$ in the 
exponential of Eq.\ (\ref{phiratio}) relative to the corresponding relation  
in Ref.\ \cite{Laenen:2000ij}, because we are computing an amplitude
here rather than a parton distribution.} mirrors the results obtained in 
Ref.~\cite{Laenen:2000ij} for the complete parton distribution.
At one loop, $B_\delta^{[q]} (\as) = (3/4) \, C_F \, (\as/\pi)$ for quarks,
and $B_\delta^{[g]} (\as) = (b_0/4) \, (\as/\pi)$ for gluons. 
From Eq.\ (\ref{phieikvirt}),  the eikonal vertex 
$\overline{\Gamma}^{\rm eik}$ and ${\cal S}$ are to be 
computed in the same fashion in perturbation theory, identifying 
their eikonal velocities with momenta 
$(\mu/\sqrt{2})\beta_p$ and $(\mu/\sqrt{2})\beta$, for incoming
and outgoing lines, respectively.
Consistency in the factorization formula Eq.\ (\ref{factorize})
then requires the same treatment of the velocity $\beta_p$
in the function ${\cal J}$ on the left-hand side of the
first equality in Eq.\ (\ref{phiratio}).

The result in \eq{phiratio} expresses the purely collinear single poles of the 
ratio between the partonic jet function and its eikonal counterpart in terms
of the virtual contribution $B_\delta (\as)$ to the splitting kernel of the
appropriate parton flavor. We now use this result in conjunction with our 
basic factorization formula, \eq{factorize}, to get another simple expression
for the function $G$.

\subsection{Relating Form Factors to Collinear Evolution Kernels}
\label{ecpfff}

We proceed by using \eq{Jpoles}, followed by \eq{phiratio},  in 
\eq{factorize} for each jet,  and exploiting renormalization-group 
invariance of the full form factor to set $\mu^2 = Q^2$.
This expresses $\Gamma(Q^2, \e)$ in terms of the eikonal soft function 
${\cal S}$, the virtual evolution kernel $B_\delta$, and the {\it finite}
factors in the partonic jet functions, as
\beqa
  \Gamma \left( 1, \as(Q^2), \e \right) & = &
  C \left( 1, \frac{(\beta_i \cdot n_i)^2}{n_i^2}, 
  \as (Q^2), \e \right) \, 
  \biggl[ \prod_{i = 1}^2
  C_J\left( \frac{(\beta_i \cdot n_i)^2}{n_i^2}, \as (\mu^2), 
  \e \right) \biggr]^{-1}
  \nonumber \\ 
  && \times \, \exp \left[ \int_0^{Q^2} \frac{d \xi^2}{\xi^2} \,
  B_\delta \left( \overline{\alpha} (\xi^2, \e) \right) \right] \,
  {\cal S} \left( \beta_1 \cdot \beta_2, \as (Q^2), \e \right) \, ,
  \nonumber \\  &\equiv&
  \overline{C} \left( \as (Q^2), \e \right) 
\label{factorizeCbar} \\ 
  && \times \, \exp \left[ \int_0^{Q^2} \frac{d \xi^2}{\xi^2} \,
  B_\delta \left(\overline{\alpha}(\xi^2, \e) \right) \right] \, 
  {\cal S} \left( \beta_1\cdot \beta_2, \as (Q^2), \e \right) \, ,
\nonumber
\eeqa
where in the second relation we define the function $\overline{C}(\as,\e)$ 
to include all factors that are finite at vanishing $\e$. 
In this function, all dependence on jet directions cancels.

In \eq{soleik}, we have an exponentiated form for the soft function ${\cal S}$, 
but to make direct contact with the standard form factor notation, \eq{sol2}, 
in terms of $G$ and $\gamma_K$, we need an expression for $\overline{C}$ 
as well.  This expression can be simply obtained, using again the 
integrability of the coupling $\as$ for $\e<0$. We write
\beqa 
  \overline{C} \left( \frac{(\beta_i \cdot n_i)^2}{n_i^2}, 
  \as (Q^2), \e \right) & = & \exp \left[ {1\over2}
  \, \int_0^{Q^2} \frac{d \xi^2}{\xi^2} 
  \, \left\{ 2 \, \xi^2 \frac{d}{d \xi^2} \,
  \log \overline{C} \left( \frac{(\beta_i \cdot n_i)^2}{n_i^2}, 
  \overline{\alpha} \left(\xi^2, \e \right), \e \right) \, \right\} \, \right]
  \nonumber \\ & \equiv &
  \exp \left[ \frac{1}{2} \, \int_0^{Q^2} \frac{d \xi^2}{\xi^2} \, 
  G_{\overline{C}} \left( \overline{\alpha} \left(\xi^2, \e \right), \e 
  \right) \, \right] \, .
\label{GbarCdef}
\eeqa
In the definition of $G_{\overline{C}}$ we have used the independence of 
the form factor on the choice of the eikonal vectors $n_i$, and 
have inserted a factor of $1/2$ to conform with the normalization of \eq{sol}. 
We note that although $G_{\overline{C}}$ is finite at $\e = 0$, the integral 
in \eq{GbarCdef} produces no poles, since the logarithmic derivative with 
respect to the scale generates a positive power of $\e$. Inserting 
\eq{GbarCdef} and \eq{soleik} in \eq{factorizeCbar}, we find our final
exponentiated result,
\beqa
  \Gamma \left( \frac{Q^2}{\mu^2}, \as(\mu^2), \e \right) & = &
  \exp \Bigg\{ \int_0^{Q^2} \frac{d \xi^2}{\xi^2} \,
  \Bigg[ G_{\rm eik} \left(1,\overline{\alpha}(\xi^2, \e) \right) + 
  2 \, B_\delta \left(\overline{\alpha}(\xi^2, \e) \right) 
  \nonumber \\ && + \, G_{\overline{C}}
  \left(\overline{\alpha}(\xi^2, \e), \e \right) -
  \frac{1}{2} \, \gamma_K \left(\overline{\alpha} (\xi^2, \e) \right) \,
  \log \left( \frac{Q^2}{\xi^2}\right) \bigg] \bigg\}\, ,
\label{fullexp}
\eeqa
where we have set $\beta_1 \cdot \beta_2 = 1$ in $G_{\rm eik}$.
Comparing this result with Eq.\ (\ref{sol2}), we conclude that
\vspace{1mm}
\beq
  G (1, \as, \e) = 
   2 B_\delta \left( \as \right)  + G_{\rm eik} \left( 1, \as \right) + 
   G_{\overline{C}} \left( \as, \e \right)~,
\label{finG}
\vspace{1mm}
\eeq
The function $G$ is thus the sum of three terms: twice the coefficient of
$\delta(1 - x)$ in the relevant parton splitting function, the
single-logarithmic anomalous dimension of the eikonal form factor, and 
finally a term associated with the running of the coupling in the 
infrared-finite hard-scattering function.  The latter term is proportional to the 
$d$-dimensional beta function, Eq.\ (\ref{dbet}) (see Eq.\ (\ref{GbarCdef})
and the Appendix); hence it vanishes as $\e\to0$ in a scale-invariant theory.

Finally, comparing Eqs.\ (\ref{Gsum}) and (\ref{finG}) for $G(\as)$,
and referring to \eq{Jderiv} and \eq{calJderiv}, which relate ${\cal G}$ to 
nonlocal matrix elements involving the field strength, we find that the
moment-independent term in the evolution kernel is determined
by the same matrix elements of the field strength, and by a combination
of anomalous dimensions of eikonal and local operators, including the
logarithmic derivative of the functions $C_J$ introduced in 
\eq{factor_parton_oe}. Explicitly,
\beqa
  B_\delta \left( \as \right) & = & \frac{1}{2} \, \sum_{i = 1}^2 \biggl[
  {\cal G}_i + \beta(\as)\frac{\partial}{\partial\as} \log C_{J_i} \biggr]
  - \frac{1}{2} G_{\rm eik} - \frac{1}{2} \gamma_{\bar{\cal S}} - 
  \gamma_J \,.
\label{Breln}
\eeqa
Although individual terms in this new relation for the function $B_\delta (\as)$
depend on $n_i$ and $\beta_i$, this dependence cancels in the sum.

Once again, \eq{finG} can easily be tested at one loop using the results 
of \secn{fafac}. It can be tested further, up to the three-loop level, by 
comparing with the results of Ref.~\cite{Moch:2005tm}, where the relation
between the function $G$ and the virtual splitting kernel was emphasized.
Indeed, Eq.~(20) of Ref.~\cite{Moch:2005tm} shows that up to three loops the 
perturbative coefficients $G^{(k)}$ of the function $G$ are
given by the sum of `maximally non-abelian' terms $f^{(k)}$, plus
twice the virtual splitting kernel contributions $B_\delta^{(k)}$, as in \eq{finG}, 
plus remainders proportional either to $\e$ or to the $\beta$ function 
coefficients. In the Appendix we show that such terms are precisely the ones 
that arise in the expansion of a function of the coupling that is defined as a 
total derivative with respect to the scale, as is the case for $G_{\overline{C}}$. 
It is natural then to identify $f^{(k)}$ with the corresponding perturbative
coefficient of $G_{\rm eik}$. Indeed, $f^{(1)} = 0$, consistently with 
\eq{geik1} in the $\MS$ scheme and for space-like kinematics. 
Furthermore, one easily verifies that, when brought to the same 
normalization, $f^{(2)}$ is one half of $\Gamma_{\rm eik}^{(2)} \equiv 
\Gamma_{\rm DY}^{(2)}$, as computed in Ref.~\cite{Belitsky:1998tc}.
The factor of $1/2$ is expected, since $f^{(2)}$ contributes to an
amplitude while $\Gamma_{\rm DY}^{(2)}$ contributes to a cross section.

Equations similar to \eq{finG} have appeared in the description
of the anomalous dimensions of effective currents (or their matching
coefficients) in soft collinear effective theory.  In Ref.~\cite{Idilbi:2006dg} 
it was noticed that the empirical relations found through three 
loops~\cite{Laenen:2005uz,Ravindran:2004mb,Moch:2005tm} imply that 
the subleading-logarithmic part of the effective-current anomalous dimension,
denoted there $B_1(\as)$, is given just in terms of $B_\delta (\as)$ and 
the function $f$ of Ref.~\cite{Moch:2005tm}; the $G_{\overline{C}}$ 
terms drop out of the anomalous dimension. In Ref.~\cite{Becher:2007ty} 
a similar relation was found, and $f$ (called $\gamma^W$ in 
Ref.~\cite{Becher:2007ty}) was identified with the anomalous 
dimension for a momentum-space Wilson loop associated with the
Drell-Yan process.  
Analogous relations were obtained for the Drell-Yan
process in Ref.~\cite{Friot:2007fd} and for deep-inelastic
scattering in Refs.~\cite{Becher:2006mr,Gardi:2007ma}.

The identification of the coefficients $f^{(k)}$ with the eikonal
(Wilson-line) quantities $G_{\rm eik}^{(k)}$ neatly explains a couple
of their properties found empirically through three 
loops~\cite{Ravindran:2004mb,Moch:2005tm}: 
the relation $f_g^{(k)}/C_A = f_q^{(k)}/C_F$,
and the maximally non-abelian color structure of these quantities.
The non-abelian exponentiation theorem for eikonal 
graphs~\cite{Gatheral:1983cz,Frenkel:1984pz} implies that only
``color-connected'' graphs composed of single gluon webs (along
with fermion-loop insertions) contribute to $G_{\rm eik}$.
Through $k=3$ loops, all such graphs have color factors of the 
``maximally non-abelian'' form $C_i C_A^{k-l-1} n_f^l$,
where $C_i$ is the Casimir factor for the eikonal line, $C_F$ for 
quarks and $C_A$ for gluons. This form breaks down at four loops, due 
to the existence of color factors that cannot be expressed in terms of 
quadratic Casimir operators, as in the case of the four-loop beta
function in QCD~\cite{van Ritbergen:1997va}.

As mentioned above, in a conformal theory there is no
contribution to \eq{finG} from $G_{\overline{C}}$ as $\e\to0$,
so we have
\beq
  G(1, \as, 0) = G_{\rm eik} \left( 1, \as \right)
  + 2  B_\delta \left( \as \right) \,.
\label{finGconformal}
\eeq
The eikonal quantity $G_{\rm eik}$ carries no information about
the spin of the parton, only its color (representation under the gauge
group).  Thus the spin-dependence of $G (1, \as, 0)$ is all carried
by the virtual part of the splitting kernel, $B_\delta (\as)$. Many conformal 
theories are supersymmetric: in this case, if two partonic states 
belong to the same supersymmetry multiplet, then they 
are in the same gauge-group representation, and $G_{\rm eik}$
is the same for both.   The leading-twist operators whose anomalous
dimensions yield $B_\delta (\as)$ will also be related by supersymmetry.
By \eq{finGconformal}, the values of $G(1, \as, 0)$ for these states
should be the same too.  This result can also be seen via supersymmetry 
Ward identities which relate the appropriate $S$-matrix elements~\cite{SWI}, 
and thereby imply that the corresponding single $1/\e$ poles have 
to be identical.  It would be interesting to see if the simple compound 
representation of $G(1,\as,0)$ in \eq{finGconformal} can help 
in the study of its properties in ${\cal N} = 4$ super-Yang-Mills theory.

\section{Concluding Remarks}
\label{concl}

We have reviewed the resummation of poles in dimensionally-regularized
singlet form factors for QCD and related theories, using the factorization 
properties of their amplitudes.  Revisiting the basic evolution equation and
its solution, we observed the strikingly simple connection between the 
analytic continuation of the form factor in a conformal theory and the
cusp anomalous dimension, given by \eq{exact}.

In view of of the universal nature of collinear poles in 
dimensionally-regularized amplitudes, \eq{amp}, we have investigated 
the origin of subleading poles in the form factor.  Our analysis extends 
the familiar relationship between the leading poles and the cusp 
anomalous dimension. We have determined the origin of the non-singular
function $G(\alpha_s)$ in the basic evolution equation, \eq{eveq}.
Relying on the operator content of the factorized jet and soft functions,
we found  \eq{Gsum}, which relates  $G(\alpha_s)$ to the short-distance
function $C$, and to $\gamma_{\bar{\cal S}}$ and $\gamma_J$, the 
anomalous dimensions of soft and jet functions respectively, as well as
to matrix elements involving the field strength and Wilson lines. The 
structure of this result is made more transparent in \eq{finG}, which 
expresses $G(\alpha_s)$  in terms of two universal quantities (the single-pole 
anomalous dimension of the eikonal form factor, and the coefficient of 
$\delta(1 - z)$ in the diagonal partonic evolution kernel), plus a 
process-dependent short-distance contribution, which generates no 
singularities and is proportional to $\epsilon$ in the conformal 
limit. The same pattern has been noted in deep-inelastic 
scattering and the Drell-Yan 
process~\cite{Laenen:2005uz,Idilbi:2006dg,Becher:2006mr,Becher:2007ty}, 
and, through the relation between form factors and
amplitudes, it will appear in subleading logarithmic corrections to any
threshold-resummed jet cross section \cite{Kidonakis:1998bk}.
Equating our two expressions for  $G(\alpha_s)$ provides an interesting 
new relation between $B_\delta (\alpha_s)$, other anomalous dimensions, 
and matrix elements of the field strength.

In summary, unlike leading poles, nonleading poles in form factors and 
fixed-angle scattering amplitudes have a compound structure, even in 
conformal theories.  Nevertheless, all contributions to $G(\alpha_s)$ 
that generate infrared poles have a well-defined and universal origin, in 
terms of matrix elements and anomalous dimensions in the massless 
gauge field theory.  In the context of planar ${\cal N} = 4$ super-Yang-Mills 
theory, the explicit operator interpretations for  $G(\as)$ may aid efforts 
to use integrability to determine its exact coupling-constant dependence.

\vspace{1cm}

\noindent
{\large \bf Acknowledgments}

\vskip0.4cm

We thank T. Becher, Z. Bern, J. Maldacena, S. Moch and M. Staudacher for
stimulating conversations, and E.\ Gardi and G.\ Grunberg for
a timely exchange relevant to integrals
over the running coupling. L.M. thanks the C.N. Yang Institute 
for Theoretical Physics at SUNY Stony Brook and the CERN Theory 
Division for hospitality during the completion of this work.
This work was supported in part by the US Department of 
Energy under contract DE--AC02--76SF00515, 
by the National Science Foundation, 
grants PHY-0354776, PHY-0354822 and PHY-0653342, 
by MIUR under contract 2006020509$\_$004, 
and by the European Community's Marie-Curie Research Training Network 
`Tools and Precision Calculations for Physics Discoveries at Colliders' 
(`HEPTOOLS'), under contract MRTN-CT-2006-035505.

\vspace{1cm}

\appendix

\section{Comparison to Fixed-Order}

The form factors for quarks and gluons in QCD have been
evaluated recently at three loops in Refs.~\cite{Moch:2005tm,Moch:2005id},
allowing for a stringent test of our result, \eq{finG}. 
Ref.~\cite{Moch:2005tm} (MVV below), in particular, gives an explicit 
expression for the single-logarithmic function $G (\as, \e)$ up to three 
loops, as a sum essentially in the form given in \eq{finG} above.  
In this sum, the function  that we denote by $G_{\overline{C}} (\as, \e)$
corresponds to a set of terms containing the coefficients
$\widetilde{G}^p_i$ in Eq.~(20) of MVV.
Similarly, $G_{\rm eik}$ can be identified with terms labeled $f$ in MVV,
whose universality is noted, without an explicit discussion of
their origin.

The key feature of Eq.~(20) in MVV is that the terms in the function $G$ 
that are not accounted for by the virtual splitting kernel $B_\delta$ or by 
the eikonal function $f$ are proportional either to $\e$, or to the coefficients 
of the $\beta$ function. We want to verify that all these terms are precisely 
of the form that follows by requiring that they are coefficients in the 
expansion of a total derivative with respect to the scale of  the running 
coupling, as is the case for our function 
$G_{\overline{C}} (\as,\e)$, \eq{GbarCdef}.

Let us begin by working out the consequences of our definition of
$G_{\overline{C}} (\as,\e)$, which is of the form
\beq
  G_{\overline{C}} \left( \as(\mu^2), \e \right) = 2 \, \mu^2 
  \frac{d}{d \mu^2} \, E \left( \as(\mu^2), \e \right) \, ,
\label{GCtotal}
\eeq
for some function $E(\as(\mu^2),\e)$ that is an expansion
in both $\as (\mu^2)$ and $\e$, and which is finite at $\e = 0$.
In terms of our factorization analysis, $E(\as,\e)$ is simply the 
logarithm of the finite coefficient function $\overline{C}$
in the factorized form 
factor, in the scheme in which the coefficient function is defined to 
absorb all finite terms in the expansion of the partonic jet functions.
Expanding $G_{\overline{C}}$ and $E$ in powers of $\as$ and $\e$,
we write
\beq
  G_{\overline{C}} \left( \as, \e \right) = \sum_{n = 1}^\infty
  G_{\overline{C}}^{(n)} (\e) \left( \frac{\as}{\pi} \right)^n =
  \sum_{n = 1}^\infty \sum_{m = 0}^\infty
  G_{\overline{C}, \, m}^{(n)} \, \e^m \, \left( \frac{\as}{\pi} \right)^n \, , 
\label{expaGE}
\eeq
and similarly for $E$. Next, we use the fact that $E$ depends on the scale 
only through the coupling $\as (\mu^2)$.  As a consequence, we can easily 
find an expression for the perturbative coefficients of $G_{\overline{C}}$ in 
terms of those of $E$. Using our normalizations for the $\beta$ function, 
Eqs.~(\ref{dbet}) and ~(\ref{beta}), we find
\beq
  G_{\overline{C}}^{(n)} (\e) = - 2 \, n \, \e \, E^{(n)} (\e) - \frac{1}{2} \,
  \sum_{k = 1}^{n -1} \, k \, b_{n - k - 1} \, E^{(k)} (\e) \, .
\label{GfromE}
\eeq
Since all $E^{(k)} (\e)$ are finite as $\e \to 0$, this implies
\beq
  G_{\overline{C}}^{(n)} (0) = - \frac{1}{2} \,
  \sum_{k = 1}^{n -1} \, k \, b_{n - k - 1} \, E^{(k)} (0) \, .
\label{G0fromE0}
\eeq
We observe from \eq{GfromE} that the perturbative coefficients
of $G_{\overline{C}}$
are sums of terms that are proportional either to $\e$ or to the 
coefficients of the $\beta$ function, as expected. Furthermore, combining 
\eq{G0fromE0} with \eq{GfromE}, it is clear that one can determine 
$G_{\overline{C}}^{(n)} (0)$ recursively in terms of $G_{\overline{C}}^{(k)} 
(\e)$, with $ k < n$. To match the notation of MVV, we proceed by defining
\beq
  \widetilde{g} (\e) \equiv \frac{1}{\e} \Big[ g(\e) - g(0) \Big] \, ,
\label{tilde}
\eeq
for any function $g (\e)$ with a finite limit as $\e \to 0$. The recursion 
starts with $G_{\overline{C}}^{(1)}(0) = 0$, so that \eq{finG} gives
\beq
  G^{(1)} = 2 B_\delta^{(1)} + G_{\rm eik}^{(1)}
  + \e \widetilde{G}_{\overline{C}}^{(1)} \,,
\label{oneloopMVV}
\eeq
matching the one-loop Eq.~(20) of MVV.

Proceeding recursively, it is easy to see that at two loops one can write
\beq
  G_{\overline{C}}^{(2)} (0) = - \frac{b_0}{2} E^{(1)} (0) = \frac{b_0}{4}
  \widetilde{G}_{\overline{C}}^{(1)} (0)~.
\label{GC2}
\eeq
Using $G_{\overline{C}}^{(2)} (\e) = G_{\overline{C}}^{(2)} (0) + \e 
\widetilde{G}_{\overline{C}}^{(2)}$, and taking into account the 
different normalizations, \eq{GC2} matches the two-loop result in 
Eq.~(20) of MVV, 
\beq
  G^{(2)} = 2 B_\delta^{(2)} + G_{\rm eik}^{(2)} 
  + \frac{b_0}{4} \widetilde{G}_{\overline{C}}^{(1)} (0)
  + \e \widetilde{G}_{\overline{C}}^{(2)} \,.
\label{twoloopMVV}
\eeq
A short calculation yields also the three-loop expression
\beqa
  G_{\overline{C}}^{(3)} (0) & = & - b_0 \, E^{(2)} (0) - \frac{b_1}{2} \,
  E^{(1)} (0) \nonumber \\ & = & \frac{b_0}{4} \,
  \widetilde{G}_{\overline{C}}^{(2)} (0) - \frac{b_0^2}{16} \,
  \widetilde{\widetilde{G} \,}_{\overline{C}}^{(1)} (0)
  + \frac{b_1}{4} \, \widetilde{G}_{\overline{C}}^{(1)} (0)~,
\label{GC3}
\eeqa
which again matches the $\beta$-function terms in the three-loop result 
in Eq.~(20) of MVV, provided the normalizations are taken into account.

An alternative way of expressing the solution of this recursion problem, 
without making use of the subtraction in \eq{tilde}, is to expand explicitly
the coefficients $G_{\overline{C}}^{(n)} (\e)$ in powers of $\e$, as done 
in \eq{expaGE}. It is straightforward to express the solution, at any order, in 
terms of the coefficients $G_{\overline{C}, \, m}^{(n)}$. One finds
\beq
  G_{\overline{C}, \, 0}^{(n + 1)} = \frac{b_0}{4} G_{\overline{C}, \, 1}^{(n)} -
  \frac{b_0^2}{16} G_{\overline{C}, \, 2}^{(n - 1)} +
  \frac{b_1}{4} G_{\overline{C}, \, 1}^{(n - 1)} + \frac{b_2}{4} 
  G_{\overline{C}, \, 1}^{(n - 2)} - \frac{b_0 b_1}{8} 
  G_{\overline{C}, \, 2}^{(n - 2)} + \ldots \, .
\label{likeELM}
\eeq
Equation~(\ref{likeELM})
was derived in Ref.~\cite{Eynck:2003fn}, as a solution to the 
problem of finding a function $\overline{G} (\as (\mu^2))$, independent of 
$\e$, but capable, upon integration over the scale, of matching the simple 
poles of the form factor, generated by the function $G(\as (\mu^2), \e)$.
The result of Ref.~\cite{Eynck:2003fn} can now be rephrased  by stating
that the function $\overline{G} (\as (\mu^2))$ must be given by 
$G (\as (\mu^2), \e = 0)$ plus the total derivative with respect to the 
scale of a finite function of $\e$.

\end{document}